\documentclass{aa}

\usepackage{txfonts}
\usepackage{longtable}
\usepackage{rotating}
\usepackage{natbib}
\usepackage{graphicx}
\usepackage{graphics}
\usepackage{psfrag}
\usepackage{amssymb}
\usepackage{xspace}
\usepackage{color}
\usepackage{enumerate}

\usepackage{array}
\newcolumntype{M}[1]{>{\centering\arraybackslash}m{#1}}
\newcolumntype{N}{@{}m{0pt}@{}}

\bibpunct{(}{)}{;}{a}{}{,}

\def\Reff{$R_{\rm eff}$}

\def\qxuv{Q_{\rm XUV}}

\def\hhh{{\rm H_3^+}}
\def\qhhh{Q_{\rm H_3^+}}
\def\hh{{\rm H_2}}
\def\hp{{\rm H^+}}
\def\hhp{{\rm H_2^+}}
\def\roche{$R_{\rm roche}$}
\def\nh{n_{\rm H}}
\def\nhh{n_{\rm H_2}}
\def\nhhh{n_{\rm H_3^+}}
\def\nhp{n_{\rm H^+}}
\def\nhhp{n_{H_2^+}}
\def\ne{n_{\rm e}}
\def\ergscm{$\rm erg\,cm^{-2}\,s^{-1}$}
\def\Teq{$T_{\rm eq}$}
\def\Teff{$T_{\rm eff}$}
\def\Rpl{$R_{\rm pl}$}
\def\Mpl{$M_{\rm pl}$}
\def\Re{\ensuremath{R_{\oplus}}}
\def\Me{\ensuremath{M_{\oplus}}}
\def\Mo{\ensuremath{M_{\odot}}}
\def\Ro{\ensuremath{R_{\odot}}}
\def\gs{$\rm g\,s^{-1}$}
\def\kms{$\rm km\,s^{-1}$}

\begin{document}
\title{A grid of upper atmosphere models for 1--40\,\Me\ planets: application to CoRoT-7\,b and HD219134\,b,c}
\subtitle{}
\author{D. Kubyshkina\inst{1}   \and
    L. Fossati\inst{1}          \and
    N. V. Erkaev\inst{2,3}      \and
    C. P. Johnstone\inst{4}     \and
    P. E. Cubillos\inst{1}      \and
    K. G. Kislyakova\inst{4,1}  \and
    H. Lammer\inst{1}           \and
    M. Lendl\inst{1}      \and
    P. Odert\inst{1,5}
}
\institute{
    Space Research Institute, Austrian Academy of Sciences, Schmiedlstrasse          6, A-8042 Graz, Austria\\
    \email{daria.kubyshkina@oeaw.ac.at}
    \and
    Institute of Computational Modelling of the Siberian Branch of the Russian Academy of Sciences, 660036 Krasnoyarsk, Russian Federation
    \and
    Siberian Federal University, 660041, Krasnoyarsk, Russian Federation
    \and
    Institute for Astronomy, University of Vienna, T\"urkenschanzstrasse 17,    A-1180 Vienna, Austria
    \and
    Institute of Physics/IGAM, University of Graz, Universit\"{a}tsplatz 5, A-8010 Graz, Austria
}
\date{}
\abstract {There is growing observational and theoretical evidence
suggesting that atmospheric escape is a key driver of planetary
evolution. Commonly, planetary evolution models employ simple
analytic formulae (e.g., energy limited escape) that are often
inaccurate, and more detailed physical models of atmospheric loss
usually only give snapshots of an atmosphere's structure and are
difficult to use for evolutionary studies. To overcome this
problem, we upgrade and employ an already existing upper
atmosphere hydrodynamic code to produce a large grid of about 7000
models covering planets with masses 1 -- 39\,\Me\ with
hydrogen-dominated atmospheres and orbiting late-type stars. The
modelled planets have equilibrium temperatures ranging between 300
and 2000\,K. For each considered stellar mass, we account for
three different values of the high-energy stellar flux (i.e., low,
moderate, and high activity). For each computed model, we derive
the atmospheric temperature, number density, bulk velocity, X-ray
and EUV (XUV) volume heating rates, and abundance of the
considered species as a function of distance from the planetary
center. From these quantities, we estimate the positions of the
maximum dissociation and ionisation, the mass-loss rate, and the
effective radius of the XUV absorption. We show that our results
are in good agreement with previously published studies employing
similar codes. We further present an interpolation routine capable
to extract the modelling output parameters for any planet lying
within the grid boundaries. We use the grid to identify the
connection between the system parameters and the resulting
atmospheric properties. We finally apply the grid and the
interpolation routine to estimate atmospheric evolutionary tracks
for the close-in, high-density planets CoRoT-7\,b and
HD219134\,b,c. Assuming the planets ever accreted primary,
hydrogen-dominated atmospheres, we find that the three planets
must have lost them within a few Myr.}
\titlerunning{A grid of upper atmosphere models for 1--40\,\Me\ planets}
\authorrunning{D. Kubyshkina et al.}
\maketitle
\section{Introduction}\label{sec:introduction}
The results of the NASA {\it Kepler} mission have revealed the presence of a large variety of planetary systems, with structures and geometries often very different from the Solar System. The detection of a large number of extra-solar planets (hereafter exoplanets) with masses and radii in between those of Earth and Neptune is a striking example \citep[e.g.,][]{bonfils2013,mullally2015}.

{Super-Earths and mini-Neptunes, absent} in the Solar System, are
extremely common and are easier to detect and characterise
compared to Earth-mass planets. Therefore, these planets are
raising great interest and are among the primary targets for
planet-finding and -characterisation missions such as CHEOPS
\citep{broeg2013}, TESS \citep{ricker2015}, CUTE
\citep{fleming2018}, JWST \citep{gardner2006,deming2009}, PLATO
\citep{rauer2014}, and ARIEL \citep{tinetti2017}.

{Super-Earths and mini-Neptunes can} have a large variety of
average densities ranging from being consistent with rocky planets
up to planets with thick hydrogen-dominated envelopes
\citep[e.g.,][]{weiss2014,lopez2014,howe2014,wolfgang2016,cubillos2017a}.
Assuming planets were formed inside the protoplanetary disk, thus
accreted a gaseous envelope, the rocky planets most likely lost
their primordial hydrogen-rich envelope through escape, while the
low-density planets still retain their primordial atmosphere.
\citet{fulton2017} revealed the presence of a dichotomy in the
radius distribution of the{ super-Earths and mini-Neptunes}
discovered by the {\it Kepler} mission \citep[see
also][]{vaneylen2018,fulton2018}, which \citet{owen2017} and
\citet{jin2017} interpreted as being the result of atmospheric
escape processes occurring during the first few hundred million
years following the dispersal of the protoplanetary disk
\citep[see][for an alternative explanation]{ginzburg2018}.

These works \citep[see also e.g.,][]{lundkvist2016} clearly showed
that atmospheric escape {is likely to play} a major role in
shaping the currently observed exoplanet population and
mass-radius distribution. Atmospheric escape is gaining also more
relevance in the characterisation of lower atmospheres: for
example, \citet{cubillos2017b} showed that the penetration depth
in the planetary atmosphere of the high-energy stellar radiation
(hereafter called XUV: EUV + X-ray) can be used to constrain the
lower pressure levels for the presence of clouds.

In this study, we focus on {1--40\,\Me\ planets that} have
accreted a primordial hydrogen-dominated envelope while forming
inside the protoplanetary nebula \citep[see e.g.,][]{stokl2016}.
Once released from the protoplanetary nebula, planets experience a
short phase of extreme hydrodynamical thermal escape, caused
mostly by their high temperature and low gravity
\citep{stokl2015,owen2016b,ginzburg2016,fossati2017}. This
so-called ``boil-off'' phase is followed by a much longer one in
which the hydrodynamic atmospheric escape is driven by the
incident stellar XUV flux \citep[e.g.,][]{lammer2003}. Usually,
this type of escape is called ``blow-off'' and the atmospheric
escape rates can be estimated using the energy- or
recombination-limited formulas
\citep{watson1981,Lecavelier2004,erkaev2007,lammer2009,ehrenreich2011,salz2016,chen2016}.
Both these escape conditions are different from classical Jeans
escape, in which only the fraction of particles lying close to or
above the exobase with velocities larger than the planetary escape
velocity leave the planet.

Overall, the energy-limited formula reproduces well the escape
rates obtained through detailed hydrodynamic upper atmosphere
modelling, particularly for close-in gas giants with atmospheres
in blow-off
\citep[e.g.,][]{lammer2009,fossati2015,salz2016,erkaev2016,erkaev2017}.
Because of its analytical form, hence allowing for fast
computations, the vast majority of planetary evolution and
population synthesis models employ the energy- and
recombination-limited formalisms to model atmospheric escape for a
wide range of planets subject to (very) different stellar
irradiation levels
\citep[e.g.,][]{jackson2012,batygin2013,jin2014,lopez2013,owen2017,jin2017,lopez2017}.
However, it has also been shown that in many cases, particularly
for highly {irradiated low-mass planets} and for planets with
hydrostatic atmospheres, the energy-limited formula tends to
significantly over- or under-estimate the escape rates
\citep[e.g.,][]{lammer2016,erkaev2015,erkaev2016,salz2016,owen2016a,fossati2017,fossati2018}.

In this work, we follow and expand on the approach of \citet{johnstone2015}, who computed a small grid of upper atmosphere hydrodynamic models and extracted the mass-loss rates by interpolating between the grid cells to model the possible
evolution of the atmosphere of early-Earth and to avoid the assumptions connected with the use of analytical formalisms. This approach enables more reliable planetary evolution computations, appropriately accounting for boil-off, blow-off, and Jeans escape, and smoothly transitioning among the different escape regimes, without significantly affecting the computational time.

We present here a large grid of upper atmosphere hydrodynamic
models computed for a wide range of parameters for {1--40\,\Me\
planets}. We present also an interpolation routine we developed to
extract model output parameters, such as atmospheric temperatures,
velocities, densities and hydrogen species abundances, and
resulting escape rates, for any planet contained within the grid
boundaries. The model grid and interpolation routine can {quickly}
produce the results of a full hydrodynamic upper atmosphere
computation for planets covered by the grid, without the need to
actually run a model. This enables faster, yet more accurate,
interpretation and characterisation of planetary atmospheres in
comparison to the results provided by, for example, the
energy-limited formula. This has now become particularly important
to understand the mass-radius-period distribution of the large
number of planets expected to be discovered in the near future by
all-sky surveys such as TESS and PLATO
\citep{rauer2014,barclay2018}. Furthermore, a grid approach
enables accurate planetary evolution and population synthesis
computations and the thorough exploration of trends in the
characteristics of planetary upper atmospheres as a function of
system parameters.

This paper is organised as follows. In Section~\ref{sec:modelling0}, we present the hydrodynamic model used to compute the grid and a comparison to the literature, while in Section~\ref{sec:grid} we describe the grid boundaries and structure. Section~\ref{sec:results} gives an overview of the results and provides a description of the interpolation routine. Section~\ref{sec:discussion} discusses the results and presents an application of the grid to the case of the low-mass, close-in planets CoRoT-7\,b and HD219134\,b,c. In Section~\ref{sec:conclusion}, we gather our conclusions.
\section{Upper atmosphere modelling}\label{sec:modelling0}
\subsection{The hydrodynamical model}\label{sec:modelling}
The construction of a large grid requires a hydrodynamic model
satisfying two basic criteria: it has to reliably compute upper
atmosphere profiles within a short time and it has to be able to
cover a wide range of stellar, orbital, and planetary parameters.
The first point is critical because the need to cover a large
parameter space requires the computation of numerous models (i.e.,
$>$1000). These criteria {are well matched} by the one-dimensional
hydrodynamic upper atmosphere model described by
\citet{erkaev2016}, which has been successfully tested for a very
wide range of planetary systems
\citep[e.g.,][]{lammer2013,lammer2016,erkaev2013,erkaev2014,erkaev2015,erkaev2016,erkaev2017,fossati2017,cubillos2017a,cubillos2017b}.

To simplify and speed up the calculation of a large number of models in the grid, we have implemented a new computational scheme, which provides an automatic selection of an initial atmospheric profile for each planet (i.e., each point in the grid). Our hydrodynamic code includes X-ray heating and $\hhh$ cooling, which are relevant for some of the planets close to our grid boundaries. The addition of X-ray heating provides us also with a further important degree of freedom relevant for young, close-in planets, which are subject to strong blow-off \citep[e.g.,][]{kubyshkina2018}. We provide below a detailed description of the modelling scheme.

We set the lower boundary of the atmospheric profile at the photospheric radius (\Rpl), where we considered the planetary atmosphere to have a temperature equal to the equilibrium temperature \citep[\Teq; see][]{fossati2017} assuming zero Bond albedo and full energy redistribution. The upper boundary was set at the Roche radius
\begin{equation}\label{eqn:Rroche}
R_{\rm roche} = d_0\left[\frac{M_{\rm pl}}{3(M_{\rm pl}+M_*)}\right]^{1/3}\,,
\end{equation}
where \Mpl\ and $M_*$ are the planetary and stellar masses,
respectively, and $d_0$ is the orbital separation. The boundary
conditions at the upper limit were set to be free, that is {the
position at which} the radial derivatives of the computed
quantities become zero. We assume a pure hydrogen atmosphere and
that at the lower boundary the atmosphere is composed exclusively
of molecular hydrogen. Following \citet{fossati2017}, for each
planet we compute the pressure at the lower boundary of the
atmosphere assuming solar abundances.

The chemical network implemented in the code accounts for hydrogen dissociation, recombination, and ionisation. In addition, the code accounts for Ly$\alpha$ cooling, XUV heating, and $\hhh$ cooling. In the literature, the height averaged heating efficiency ($\eta$), which is the fraction of absorbed stellar XUV radiation converted into thermal energy of the atmosphere, ranges between 10\% and 60\% \citep[e.g.,][]{watson1981,yelle2004,murray2009,cecchi2009,owen2012,shematovich2014,salz2016}. \citet{salz2016} showed that for Earth- to Jupiter-mass planets $\eta$ varies approximately between 10\% and 25\%. The implementation of a self-consistent calculation of the heating efficiency would have made the hydrodynamic code too slow to allow the computation of a large grid. For this reason, we decided to follow the considerations of \citet{erkaev2016} adopting for all planets a constant $\eta$ value of 15\% at all wavelengths.

The code solves the equations for mass, momentum, and energy conservation
\begin{eqnarray}
\frac{\partial\rho}{\partial t} + \frac{\partial(\rho v r^2)}{r^2\partial r} &=& 0\,, \\
\frac{\partial\rho v}{\partial t} + \frac{\partial[r^2(\rho v^2+P)]}{r^2\partial r} &=& - \frac{\partial U}{\partial r} + \frac{2P}{r}\,, \\
\frac{\partial[\frac{1}{2}\rho v^2+E+\rho U]}{\partial t} &+& \frac{\partial vr^2[\frac{1}{2}\rho v^2+E+P+\rho U]}{r^2\partial r} = \nonumber\\
Q_{\rm XUV} - Q_{\rm Ly\alpha} &+& \frac{\partial}{r^2\partial r}(r^2\chi \frac{\partial T}{\partial r}) - Q_{\rm H_3^+}\,,
\end{eqnarray}
where $\rho$, $v$, and $T$ are the mass density, bulk velocity, and temperature as a functions of the radial distance from the planetary center $r$, respectively. The quantity
\begin{equation}
\label{eq:gravpot} \mathbf{U = U_0\left[-\frac{1}{\zeta}-\frac{1}{\delta(\lambda-\zeta)}-\frac{1+\delta}{2\delta\lambda^3}\left(\lambda\frac{1}{1+\delta}-\zeta\right)^2\right]}
\end{equation}
%
is the planetary gravitational potential accounting for the Roche lobe effect \citep{erkaev2007}. In Equation~(\ref{eq:gravpot}), $U_0 = GM_{\rm pl}/R_{\rm pl}$, $\delta = M_{\rm pl}/M_*$, $\lambda = d_0/R_{\rm pl}$ (where $d_0$ is the orbital separation), and $\zeta = R/R_{\rm pl}$. The term
\begin{equation}
\chi = 4.45\times10^4\left(\frac{T[{\rm K}]}{1000}\right)^{0.7}\,,
\end{equation}
in erg\,cm$^{-1}$\,s$^{-1}$, is the thermal conductivity of the
neutral gas \citep{watson1981}. {\bf We do not account for the
thermal conductivity of electrons and ions. We tested this
assumption concluding that the inclusion of this effect leads to
negligible variations to the results. In particular, the largest
effect is on the temperature maximum for highly irradiated,
high-density planets, which decreases by as much as $7\%$, with a
typical decrease lying around $1-2\%$.}

The terms $P$ and $E$ are the atmospheric pressure and thermal
energy, which are defined as
\begin{equation}
P = (n_{\rm H} + n_{\rm H^+} + n_{\rm H_2} + n_{\rm H_2^+} + n_{\rm H_3^+}+n_{\rm e})kT
\end{equation}
and
\begin{equation}
E = \left[\frac{3}{2}(n_{\rm H}+n_{\rm H^+}+n_{\rm e})+\frac{5}{2}(n_{\rm H_2}+n_{\rm H_2^+})+3\,n_{\rm H_3^+}\right]kT\,.
\end{equation}
Finally, $Q_{\rm XUV}$, $Q_{\rm Ly\alpha}$, and $Q_{\rm H_3^+}$
are the volume heating/cooling  rates for XUV heating, $\rm
Ly\alpha$ cooling, and $\hhh$ cooling, respectively.

The spectral dependence of the stellar XUV flux varies significantly from star to star. Since we aim at computing a grid of models valid for a wide range of system parameters, it is impossible to account for the full spectral dependence of the stellar XUV flux, though the code would in principle allow it. For this reason, we assumed that the whole stellar EUV flux is emitted at a single wavelength of 60\,nm \citep{murray2009}. To account for X-ray heating, we assumed that the stellar X-ray photons are all emitted at a wavelength of 5\,nm, roughly in the middle of the X-ray wavelength band.

The XUV heating function $\qxuv$ is therefore composed of two terms, $Q_{\rm EUV}$ and $Q_{\rm X}$, which describe the heating by the EUV and X-ray stellar flux, respectively. These two functions are constructed in the same way, except for the absorption cross-sections and absorption functions of the stellar flux inside the planetary atmosphere that are defined at 5 and 60\,nm. The total heating function thus becomes $\qxuv$\,=\,$Q_{\rm EUV} + Q_{\rm X}$. Each heating function takes the form of
\begin{equation}\label{eq:Qxuv}
Q_{m} = \eta\,\sigma_{m}\,(\nh+\nhh)\,\phi_{m}\,,
\end{equation}
where $m$ stands for either EUV or X, $\sigma_{m}$ is the absorption cross-section for the specific wavelength, and $\phi_{m}$ is the flux absorption function
\begin{equation}
\label{eq:Jm}
\phi_{m} = \frac{1}{4\pi}\int_{0}^{{\pi}/{2}+\arccos({1}/{r})}\{J_{m}(r,\theta) \times 2\pi \sin(\theta) \}\,d\theta\,.
\end{equation}
In Equation~(\ref{eq:Jm}), $J_{m}(r,\theta)$ is a function in spherical coordinates describing the spatial variation of the EUV, or X-ray, flux due to atmospheric absorption \citep{erkaev2015} and $r$, in this case, corresponds to the radial distance from the planetary center.

We defined the absorption cross-section as $\sigma = \sigma_0\left({E_{\lambda}}/{E_{i}}\right)^{-3}$, where $\sigma_0 = 6\times10^{-18}$, $E_{i}$\,=\,13.6\,eV is the hydrogen ionisation energy, and $E_{\lambda}$ is the photon energy in a specific wavelength range ($E_{\lambda}$\,=\,20\,eV in the EUV and 248\,eV in the X-ray domain). It follows that the EUV flux absorption cross-sections are $2\times10^{-18}$\,cm$^{-3}$ and $1.2\times10^{-18}$\,cm$^{-3}$ for atomic and molecular hydrogen, respectively \citep{spitzer1978}. The X-ray absorption cross-section is approximately three orders of magnitude smaller than the EUV one. This implies that the stellar X-ray photons penetrate deeper into the planetary atmosphere than the EUV photons, thus heating the atmosphere closer to \Rpl. For this reason, despite that stellar X-ray fluxes are significantly smaller than the EUV
fluxes, X-rays can still cause significant atmospheric heating.

We implemented Ly$\alpha$ cooling by adding the following function to the energy conservation equation \citep{watson1981}
\begin{equation}\label{eq:Lalpha}
Q_{\rm Ly\alpha} = 7.5 \times 10^{-19}\,n_{\rm e}\,n_{\rm
H}\,\exp\left(-\frac{118348}{T[{\rm K}]}\right)\,.
\end{equation}
To implement $\hhh$ cooling, we followed \citet{miller2013} and added in the energy conservation equation the function
\begin{equation}\label{eq:Qh3plus}
\qhhh = 4\,\pi\,\nhhh\,e^{\sum_{n} C_{n}T^n}\,,
\end{equation}
where $C_n$ are the temperature-dependent coefficients listed in
Table\,5 of \citet{miller2013}. {Heating rates in \ref{eq:Qxuv},
\ref{eq:Lalpha} and \ref{eq:Qh3plus} are given in
erg\,cm$^{-3}$\,s$^{-1}$.}

{Since we consider non-magnetic planets, we did not include
electric conduction due to ionised components. If large enough,
conduction prevents the penetration of the interplanetary magnetic
field inside the ionosphere, which results in the formation of a
magneto pause separating the stellar wind protons from the
atmospheric ions. In addition, in case of a strongly magnetised
planet, the hydrodynamic flow of the escaping ionised gas can
produce electric currents in the ionosphere, which would generate
a resisting force against the escaping hydrodynamic flow.}

The complete list of chemical reactions and the relative cross-sections ($\nu_{\rm H}$, $\nu_{\hh}$, $\alpha_{\rm H}$, $\alpha_{\hh}$, $\nu_{\rm diss}$, $\gamma_{\rm H}$, $\nu_{\rm Hcol}$, $\gamma_{\hh}$, $\alpha_{\hhh1}$, $\alpha_{\hhh2}$) considered in the model are listed in Appendix~\ref{apx:A}. The continuity equations connected with the chemical reactions are
\begin{eqnarray}\label{eqn:species}
\frac{\partial \nh}{\partial t}+\frac{\partial(\nh
vr^2)}{r^2\partial r} &=& -\nu_{\rm H}\nh -
\nu_{\rm Hcol}\ne\nh  \nonumber \\
+ \alpha_{\rm H}\ne\nh &+& 2\alpha_{\hh}\ne\nhh +
2\nu_{dis}\nhh n \nonumber \\
- 2\gamma_{\rm H}n\nh^2 &+& \gamma_{\hh}(\nhh\nhhp-\nh\nhhh) \nonumber \\
+ (\alpha_{\hhh1} &+& 3\alpha_{\hhh2})\nhhh\ne\,, \\
\frac{\partial \nhh}{\partial t}+\frac{\partial(\nhh
vr^2)}{r^2\partial r} &=&
-\nu_{\hh}\nhh - \nu_{\rm dis}\nhh n \nonumber \\
+ \gamma_{\rm H}n\nh^2 &+&
\gamma_{\hh}(\nh\nhhh - \nhh\nhhp) \nonumber \\
&+& \alpha_{\hhh1}\nhhh\ne\,, \\
\frac{\partial \nhp}{\partial t}+\frac{\partial(\nhp
vr^2)}{r^2\partial r} &=& \nu_{\rm H}\nh +
\nu_{\rm Hcol}\ne\nh \nonumber - \alpha_{\rm H}\ne\nhp\,, \\
\frac{\partial \nhhp}{\partial t}+\frac{\partial(\nhhp
vr^2)}{r^2\partial r} &=&
\nu_{\hh}\nhh - \alpha_{\hh}\ne\nhhp \nonumber \\
+ \gamma_{\hh}(\nh\nhhh &-& \nhh\nhhp)\,, \\
\frac{\partial \nhhh}{\partial t}+\frac{\partial(\nhhh
vr^2)}{r^2\partial r} &=&
\gamma_{\hh}(\nhh\nhhp-\nh\nhhh) \nonumber \\
- (\alpha_{\hhh1}&+&\alpha_{\hhh2})\nhhh\ne\,.
\end{eqnarray}
Here, the electron density is defined as
\begin{equation}
\ne = \nhp + n_{\rm H_2^+} + \nhhh\,,
\end{equation}
while the total hydrogen number density is the sum of the number density of all species. Finally, the mass density is
\begin{equation}\label{rho}
\rho = m_{\rm H}(n_{\rm H}+n_{\rm H^+}) + 2m_{\rm H}(n_{\rm H_2}+n_{\rm H_2^+}) + 3m_{\rm H}n_{\rm H_3^+}\,.
\end{equation}

For computational convenience (e.g., simplification of the continuity equations), we apply the set of normalisations presented in Appendix~\ref{apx:B}. The numerical solution is based on the finite differential McCormack scheme \citep[Predictor-Corrector-Method; see][for more details]{erkaev2016}.
\subsection{Comparison with previous results}\label{sec:comparison}
To test the modelling results, we compared the mass-loss rates obtained for a sample of previously (observationally and/or theoretically) studied planets with those present in the literature (Table~\ref{tab:comparison}). Of the four planets considered in this comparison, just GJ\,436\,b and Kepler-11\,b fall within the grid boundaries and the inclusion in the comparison of the two classical hot Jupiters, HD209458\,b and HD189733\,b, is due to the fact that these are the best studied systems in terms of atmospheric escape. For our calculations, we employed the stellar XUV fluxes and masses given by \citet{guo2016}.
\begin{table*}
\caption{Comparison between the mass-loss rates obtained from our hydrodynamic modelling (column six), from the energy-limited formula (column seven), and from the literature (column eight). The last column lists also the source of the published mass-loss rates.} \label{tab:comparison} \centering
\begin{tabular}{l|c|c|c|c|c|c|l}
\hline \hline
ID & $\Lambda$ & $d_0$ & $F_{\rm XUV}$ & $M_*$ & $\dot{M}$ & $\dot{M}_{\rm en}$ & $\dot{M}_{\rm publ}$ \\
   &           & [AU]  & [\ergscm]     & [\Mo] & [\gs]     & [\gs]              & [\gs]                \\
\hline
HD209458\,b &  90 & 0.047   & 1086  & 1.148     & 1.2$\times10^{10}$ & 8.0$\times 10^{9}$  & 3.3$\times 10^{10}$ (a) \\
            &     &         &       &           &                    &                     & $0.6-10\times 10^{10}$ (b) \\
            &     &         &       &           &                    &                     & $1.9\times 10^{10}$ (c) \\
\hline
GJ\,436\,b  & 58  & 0.02887 & 1760  & 0.452     & 3.95$\times10^{9}$  &  2.9$\times10^{9}$  & 1$\times 10^8-1\times 10^9$ (d) \\
            &     &         &       &           &                    &                     & 1$\times 10^{10}$ (e) \\
            &     &         &       &           &                    &                     & 2.2$\times 10^{10}$ (f) \\
            &     &         &       &           &                    &                     & 4.5$\times 10^9$ (c)  \\
\hline
Kepler-11\,b& 18  & 0.091   & 278   & 0.95      & 1.1$\times10^{9}$  &  7.5$\times 10^{8}$ & $1.15-2\times 10^8$ (g)\\
            &     &         &       &           &                    &                     & $1.17-1.3\times 10^7$ (h) \\
            &     &         &       &           &                    &                     & $1\times 10^9$ (e) \\
\hline
HD189733\,b & 179 & 0.03    & 24778 & 0.8       & 4.9$\times 10^{9}$ &  $4.8\times 10^{10}$& $0.04-10\times 10^{10}$ (b) \\
            &     &         &       &           &                    &                     & $5-9\times 10^{11}$ (e)\\
            &     &         &       &           &                    &                     & $4.1\times 10^9$ (c)  \\
\hline
\end{tabular}
\tablefoot{ a -- \citet{murray2009}; b -- \citet{bourrier2013}; c -- \citet{salz2016}; d -- \citet{ehrenreich2015}; e -- \citet{guo2016}; f -- \citet{bourrier2016}; g -- \citet{lammer2013}; h -- \citet{kislyakova2014}.}
\end{table*}

We find good agreement between our values and those published in
the literature, in particular for HD209458\,b, GJ\,436\,b, and
Kepler-11\,b. Note that for Kepler-11\,b \citet{kislyakova2014}
considered mostly non-thermal escape, which is significantly
smaller than the XUV driven escape, while \citet{lammer2013}
adopted a completely different lower boundary condition, which led
to a significant underestimation of the mass-loss rates
\citep[see][for more details]{lammer2016}. In case of HD189733\,b,
our estimation {lies within the interval given by}
\citet{bourrier2013}, but significantly {below that of
\citet{guo2016}, (which appears to be} an outlier compared to
other estimations), and it is an order of magnitude {smaller than
that} given by the energy-limited formula. The reason may be that
Equation~(\ref{eq:Lalpha}) possibly overestimates the cooling for
hot Jupiters, which are optically thick to L$\alpha$ in the region
where the cooling peaks, so the radiation does not escape
efficiently. This {was addressed in detail} by \citet{menager2013}
and \citet{koskinen2013}.

The works that most closely {resemble our} are those of
\citet{murray2009}, \citet{guo2016}, and \citet{salz2016}, with
which we find good agreement. We remark that none of the
comparison mass-loss rates was computed with the energy-limited
approximation.
\section{Model grid}\label{sec:grid}
We designed the grid to model {super-Earths and mini-Neptunes}
orbiting main-sequence stars. The computations were made
considering the following system parameters: planetary mass \Mpl,
planetary radius \Rpl, equilibrium temperature \Teq, orbital
separation d$_0$, stellar mass $M_*$, and the stellar XUV flux at
the planetary orbit $F_{\rm XUV} = F_{\rm EUV} + F_{\rm X}$. {As
mentioned above, we consider the} planetary radius \Rpl\ to be
equal to the photospheric radius assuming a clear
hydrogen-dominated atmosphere and solar abundances.

The stellar temperature and radius change along the main-sequence
phase of evolution, defined as in \citet{yi2001}. Consequently,
each \Teq\ value corresponds to a range of possible orbital
separations defined by the possible range of changes in stellar
parameters. By fixing stellar equilibrium temperature and radius,
this range of orbital separations corresponds to \Teq\ variations
of the order of 10--20\,K. To save computation time, we adopted
one value of the orbital separation, namely that at the center of
the range, for each \Teq\ value. Therefore, d$_0$ is derived from
the stellar mass and equilibrium temperature. {This implies that
just} five input parameters of the grid are independent.

{We computed models for planets with masses ranging between 1 and
39\,\Me\  (i.e., up to twice the mass of Neptune or one tenth of
Jupiter's), with a variable step size that increases
logarithmically with mass for a total of 14 planetary mass values.
The planetary radius ranges between 1 and 10\,\Re\  (i.e., up to
one Jupiter radius and 2.5 times Neptune's), in regular steps of
1\,\Re\  (i.e., total of 10 planetary radius values). The
equilibrium temperature of the planets in the grid ranges between
300 and 2000\,K with regular steps of 400\,K (i.e., total of 5
temperature values).} The cooler boundary was set to cover planets
orbiting in the habitable zone, while the hotter boundary was set
to ensure that our assumption on the composition of the atmosphere
at the lower boundary (i.e., H$_2$-dominated) holds
\citep{koskinen2010}. {Our focus is on planets orbiting early M-
to late F-type stars, thus we considered stellar masses between
0.4 and 1.3\,\Mo\ for a total of five different stellar masses. We
plan to extend the grid to lower mass stars, which are primary
targets for various planet-finding facilities, such as CARMENES
\citep{carmenes} and TESS \citep{ricker2015}.
Table~\ref{tab:gridpar} lists the values of stellar mass,
equilibrium temperature, planetary radius, and planetary mass
considered for the computation of the grid.}
\begin{table}
\caption{{List of stellar masses, equilibrium temperatures,
planetary radii, and planetary masses considered for the
computation of the grid.}} \label{tab:gridpar}
\begin{center}
\begin{tabular}{c|c|c|c}
\hline
\hline
$M_*$ & \Teq & \Rpl & \Mpl \\
\Mo & K & \Re & \Me \\
\hline
0.4 & 300  & 1.0  & 1.0 \\
0.6 & 700  & 2.0  & 1.6 \\
0.8 & 1100 & 3.0  & 2.1 \\
1.0 & 1500 & 4.0  & 3.2 \\
1.3 & 2000 & 5.0  & 4.3 \\
    &      & 6.0  & 5.0 \\
    &      & 7.0  & 6.7 \\
    &      & 8.0  & 7.8 \\
    &      & 9.0  & 9.0 \\
    &      & 10.0 & 12.1 \\
    &      &      & 16.2 \\
    &      &      & 21.7 \\
    &      &      & 29.1 \\
    &      &      & 39.0 \\
\hline
\end{tabular}
\end{center}
\end{table}

{We set the range of orbital separations covered by the grid on
the basis of the stellar mass and planetary equilibrium
temperature, thus stellar radius ($R_*$) and effective temperature
(\Teff). The two last quantities were derived considering the
range of radii and effective temperatures covered by a star of
each considered mass along the main-sequence on the basis of
stellar evolutionary tracks \citep{yi2001}. This leads to the fact
that only a limited range of orbital separations had to be
considered for each given stellar mass, saving computation time.
Considering all stellar masses, the orbital separation ranges
between 0.002 and 1.3\,AU.}

{For the XUV stellar fluxes, we considered three distinct values
corresponding roughly to a chromospherically active star, a
moderately active star, and a quiet star. To set the high XUV flux
value, we considered that the X-ray saturation threshold observed
for main-sequence late-type stars lies at roughly $L_{\rm
X}/L_{\rm bol} = 10^a$, where $L_{\rm X}$ is the X-ray luminosity,
$L_{\rm bol}$ is the bolometric luminosity at the zero age main
sequence \citep{yi2001}, and $a$ ranges between $-$2.5
\citep[e.g.,][]{reiners2014} and $-$3.1
\citep[e.g.,][]{wright2011}. We therefore set the maximum X-ray
luminosity as $L_{\rm Xmax}/L_{\rm bol} = 5\times10^{-3}$ and the
minimum X-ray luminosity as $L_{\rm Xmin}/L_{\rm bol} = 10^{-7}$.
The EUV stellar luminosity was then derived from the X-ray
luminosity following \citep{sanz2011}}
\begin{equation}
\log{L_{\rm EUV}} = 4.8 + 0.86\,\log{L_{\rm X}}\,.
\label{eq:X2EUV}
\end{equation}
{The specific X-ray and EUV luminosities adopted for each stellar
mass are listed in Table~\ref{tab:xuv}.}
\begin{table*}
\caption{X-ray and EUV luminosities adopted for each stellar mass. The subscripts ``1'', ``2'', and ``3'' indicate cases of inactive, moderately active, and active stars, respectively.}
\label{tab:xuv}
\begin{center}
\begin{tabular}{c|c|c|c|c|c|c|c}
\hline
\hline
$M_*$ & $L_{\rm bol}$ & $L_{\rm X1}$ & $L_{\rm X2}$ & $L_{\rm X3}$ & $L_{\rm EUV1}$ & $L_{\rm EUV2}$ & $L_{\rm EUV3}$ \\
\Mo & [$10^{31}$\,erg\,s$^{-1}$] & [$10^{24}$\,erg\,s$^{-1}$] & [$10^{26}$\,erg\,s$^{-1}$] & [$10^{27}$\,erg\,s$^{-1}$] & [$10^{25}$\,erg\,s$^{-1}$] & [$10^{27}$\,erg\,s$^{-1}$] & [$10^{28}$\,erg\,s$^{-1}$] \\
  \hline
  0.4 & $4.29$   & $4.3$   & $1.1$  & $6.2$    & $9.6$    & $1.6$  & $5.0$ \\
  0.6 & $27.82$  & $32.5$  & $1.1$  & $29.8$   & $55.5$   & $1.6$  & $19.4$ \\
  0.8 & $94.72$  & $363.9$ & $40.5$ & $102.2$  & $438.8$  & $34.9$ & $55.8$ \\
  1.0 & $266.41$ & $363.9$ & $40.5$ & $286.8$  & $438.8$  & $34.9$ & $135.7$ \\
  1.3 & $942.41$ & $942.4$ & $73.2$ & $1011.4$ & $995.1$  & $58.5$ & $402.2$ \\
  \hline
\end{tabular}
\end{center}
\end{table*}

\begin{figure}[ht!]
\includegraphics[width=\hsize,clip]{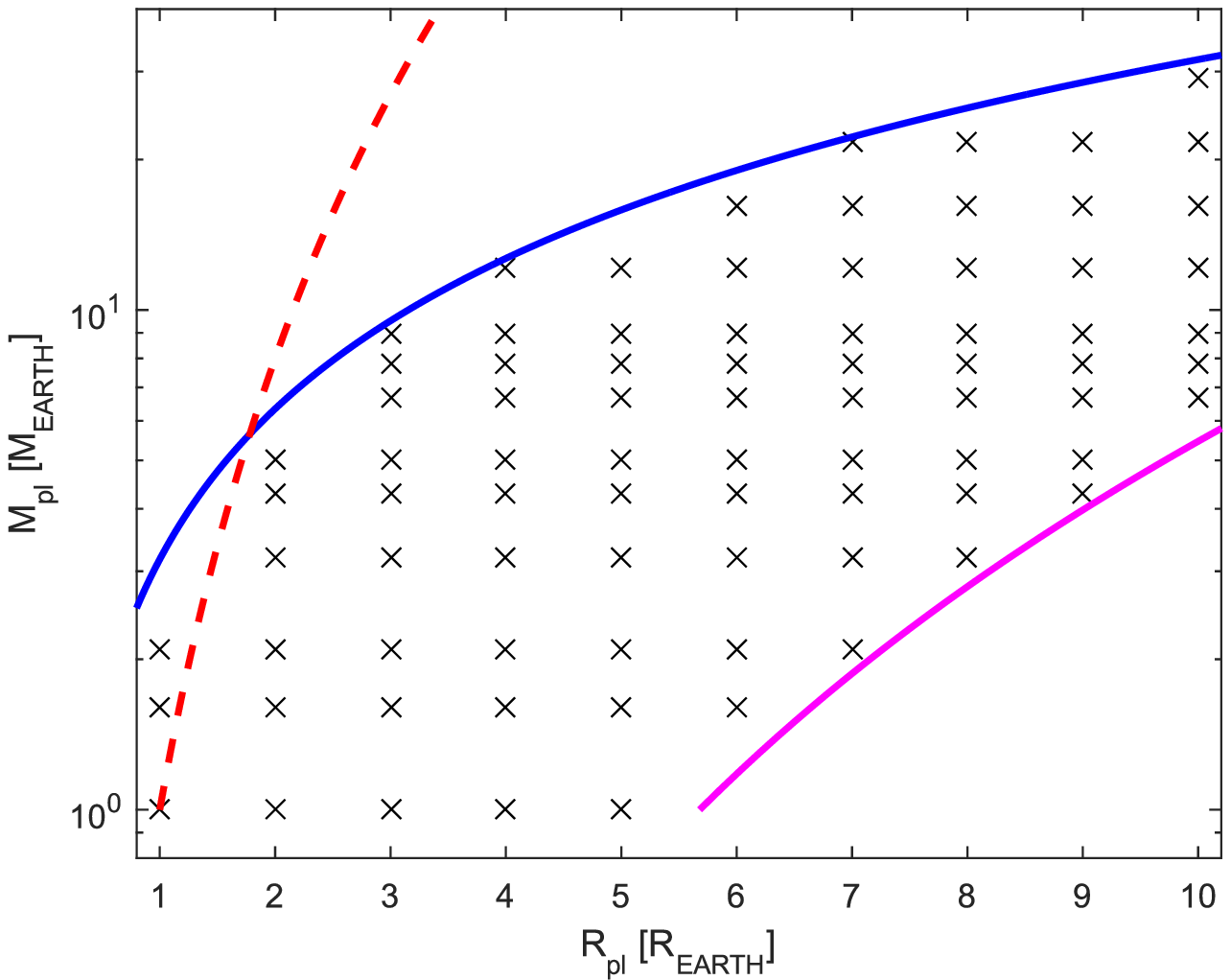}
\includegraphics[width=\hsize,clip]{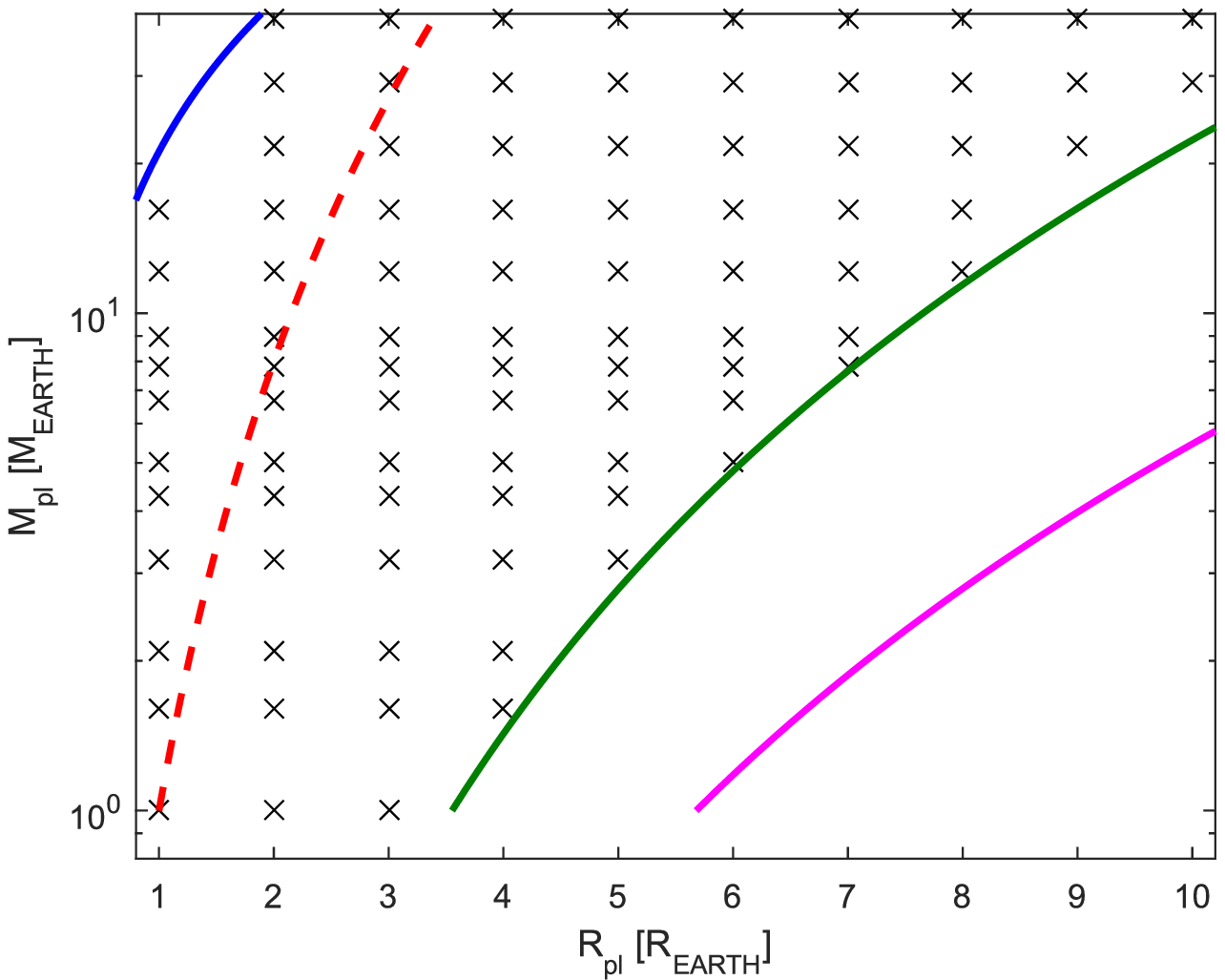}
\caption{{Position of some of the modelled planets (black crosses)
in the mass-radius diagram. All planets orbit a 1\,\Mo\ star and
have an equilibrium temperature of 300\,K (top) and 2000\,K
(bottom). The blue and magenta solid lines indicate the boundaries
of the grid set by the cut on $\Lambda$ and on bulk density,
respectively. Since $\Lambda$ depends on \Teq, the top boundary is
different in the different panels. The green solid line indicates
the boundary of the grid set by the cut on the Roche lobe. The
position of this boundary depends on the orbital separation, thus
on \Teq. For reference, the red dashed line indicates Earth's
density.}} \label{M_R}
\end{figure}

{To avoid spending time calculating planets that probably do not
exist in nature, we restricted the computations to planets with an
average density larger than 0.03\,g\,cm$^{-3}$ \citep[equal to the
lowest known measured density][]{masuda2014} and a restricted
Jeans escape parameter $\Lambda$ smaller than 80 (where
atmospheres are presumably stable), where
\citep{jeans1925,chamber1963,opik1963,fossati2017}}
\begin{equation}
\label{eq:lambda}
\Lambda = \frac{G M_{\rm pl}m_{\rm H}}{k_{\rm b}T_{\rm eq}R_{\rm
pl}}\,.
\end{equation}
{$\Lambda$ is the value of the Jeans escape parameter calculated
at the observed planetary radius and mass for the planet's
equilibrium temperature and considering atomic hydrogen,
independent of the atmospheric temperature profile. We further
excluded planets where the Roche lobe is closer than 0.5 planetary
radii from the surface. This cut is most relevant for the hottest
planets ($>$\,1500\,K) orbiting stars less massive than about
0.8\,\Mo. As an example, Figure~\ref{M_R} shows the positions of
the modelled planets in the mass-radius diagram at two different
equilibrium temperatures.}

{To summarise, our grid consists of five data points for stellar
mass and planetary equilibrium temperature, each, ten data points
for planetary radius, 14 data points for planetary mass, and three
data points for stellar XUV luminosity. This leads to a total of
10\,500 models. However, because of the restrictions described
above, the total number of models in the grid reduces to 6\,700.}
\section{Results}\label{sec:results}
For each modelled planet, we computed the main atmospheric parameters as a function of the radial distance from the planetary center. These include the atmospheric temperature, number density, bulk velocity, X-ray and EUV volume heating rates, and abundance of the considered species ($\hh$, H, $\hhp$, $\hp$, $\hhh$, e). From these quantities, we estimated the positions of the maximum dissociation and ionisation (the distances corresponding to the maximum of the number densities of atomic and ionised hydrogen, respectively), the mass-loss rate $\dot{M}$, and the effective radius of the XUV absorption that is defined as
\begin{equation}\label{eqn:reff}
R_{\rm eff} = R_{\rm pl}\sqrt{1 + 2\int_1^{\infty}\frac{J_{\rm XUV}(r,\frac{\pi}{2})}{F_{\rm XUV}}rdr}\,,
\end{equation}
where $J_{\rm XUV}(r,\frac{\pi}{2})$ is the XUV flux as it travels through the planetary atmosphere along the star-planet direction and is mostly determined by the density $n(r)$. The mass-loss rate is computed as the product of the atmospheric density and velocity at the upper boundary. To account for the fact that we employ a one dimensional model, this value is then multiplied by the surface area of a sphere with radius equal to $R_{\rm roche}$. For most planets, hydrogen dissociation occurs in a relatively narrow range of distances, which is typically smaller than one planetary radius.
\begin{figure}
\includegraphics[width=\hsize,clip]{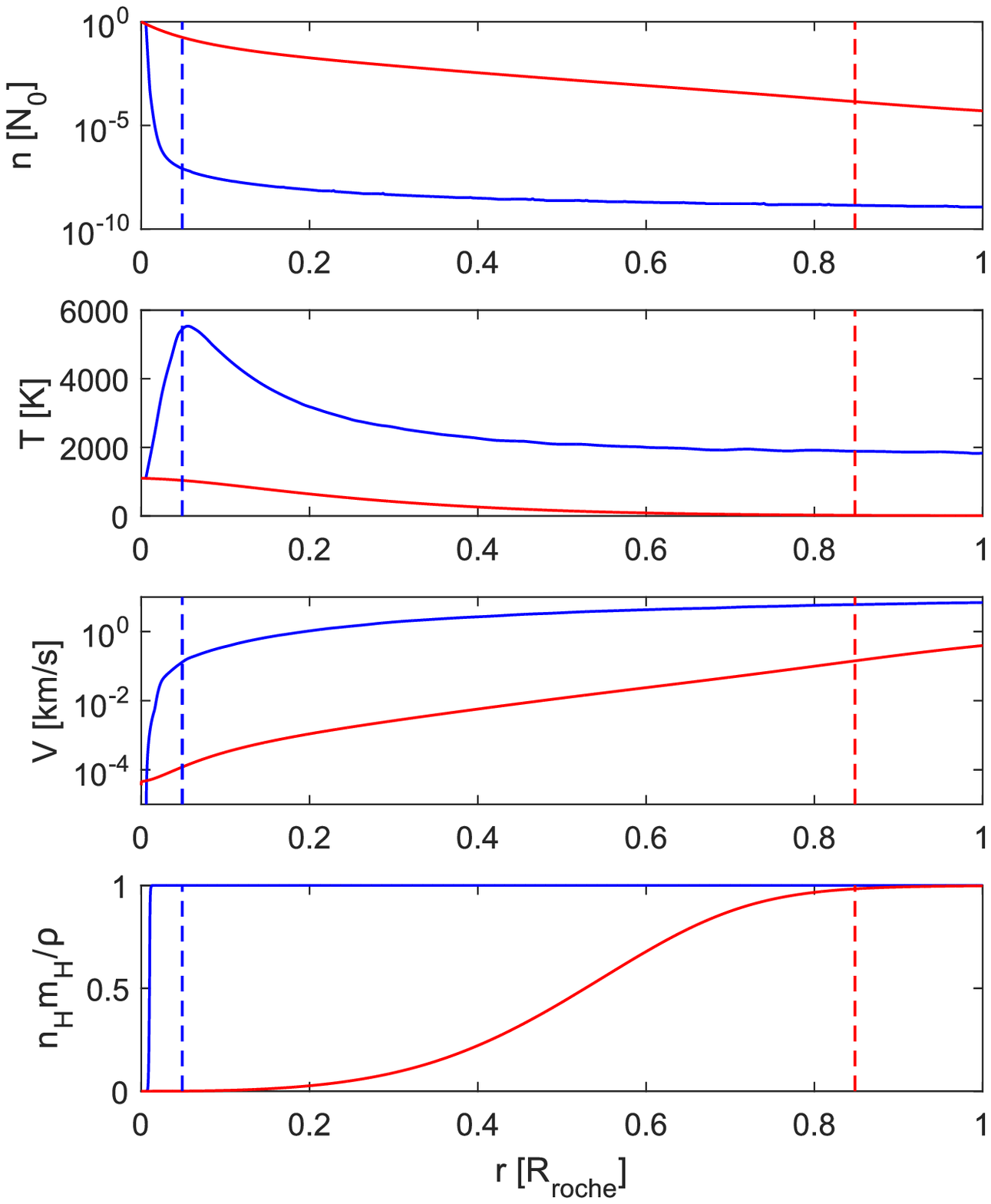}
\caption{From top to bottom: atmospheric profiles for density, temperature, velocity, and fraction of atomic hydrogen for planets with $\Lambda$ equal to 4.8 (red) and 66.7 (blue). Both planets orbit a 1\,\Mo\ star, have \Teq\,=\,1100\,K and \Rpl\,=\,3\,\Re, and are subject to an incident XUV flux of
92.6\,\ergscm. The planet with the lower $\Lambda$ has a mass of 2.1\,\Me, while the other one has a mass of 29.1\,\Me. The density (top) is normalised to its value at \Rpl. The blue and red dashed lines show the effective radii of the XUV absorption. To allow comparing planets with significantly different Roche radii, the x-axis is in units of the planetary Roche radii, starting from the planetary surface.}
\label{fig:profiles}
\end{figure}

Although the atmospheric parameters vary significantly across the
grid, there are some common characteristics. One of the most
important ones is that the atmospheric behaviour strongly depends
on $\Lambda$. For planets with low $\Lambda$ values (i.e.,
$\lesssim$10), the atmosphere is weakly bound to the planet and
experiences strong boil-off. {The energy budget of these planets
is determined by adiabatic cooling and the mass-loss rates are not
significantly affected by variations in the stellar XUV} flux.
With increasing $\Lambda$, the role of planetary gravity in the
atmospheric dynamics decreases and the atmosphere gradually
switches to being controlled by the stellar XUV heating. We find
that the border between these two regimes lies at $\Lambda$ values
ranging between 10 and 30, depending on the system parameters, in
agreement with \citet{fossati2017}.

As an example, Figure~\ref{fig:profiles} compares the atmospheric density, temperature, velocity, and atomic hydrogen abundance profiles for two planets with $\Lambda$ equal to 4.8 and 66.7. The two planets orbit a 1\,\Mo\ star, have an equilibrium temperature of 1100\,K (i.e., $d_0$\,=\,0.075\,AU), a radius of
3\,\Re, and are subject to an incident XUV flux of 92.6\,\ergscm. The planet with the lower $\Lambda$ has a mass of 2.1\,\Me, while that with the higher $\Lambda$ has a mass of 29.1\,\Me. For the less massive planet, we derived a value of the effective XUV absorption radius (\Reff) of 5.5\,\Rpl, a Roche radius of 6.5\,\Rpl, and a mass-loss rate of $1.1\times10^{14}$\,\gs. For the more massive planet, we derived a \Reff\ value of 1.2\,\Rpl, a Roche radius of 17.1\,\Rpl, and a mass-loss rate of $4.0\times10^{7}$\,\gs. For the less massive planet, we found also that the velocity of the atmospheric particles becomes supersonic close to the Roche radius (at 6.2\,\Rpl), while for the more massive planet the particles become supersonic well below the upper boundary (at 9.1\,\Rpl).

Figure~\ref{fig:profiles} shows how the density decreases with increasing distance from the planetary surface; the decrease is steeper for the more massive planet, because it hosts a more compact atmosphere (because of stronger gravity). The temperature profiles show that the stellar XUV flux efficiently heats the more massive planet inducing a temperature peak at the thermospheric
level, reaching its maximum close to \Reff, where the model indicates also the presence of strong $\hh$ dissociation. At higher altitudes, the atmosphere is composed fully of atomic hydrogen and is dominated by adiabatic cooling, which is caused by the atmosphere's expansion and dominates over XUV heating. In contrast, for the less massive planet, the XUV stellar flux does not penetrate deep enough into the planetary atmosphere to cause thermospheric heating, thus the atmosphere expands adiabatically, driven by its internal heat and by the low planetary gravity. In general, the profiles of planets with low $\Lambda$ values do not develop steep gradients \citep[see e.g.,][]{kubyshkina2018}, making the definition of \Reff\ and of the position of the maximum dissociation and ionisation ambiguous. We will come back to this point in Section~\ref{sec:barriers}.

\begin{figure*}[ht!]
\begin{rotate}{90} \hspace{2.3 cm} $d_0 = 0.0404$\,au \hspace{2.3 cm} $d_0 = 0.0751$\,au \hspace{2.3 cm} $d_0 = 0.1854$\,au \end{rotate}
\includegraphics[width=\hsize,clip]{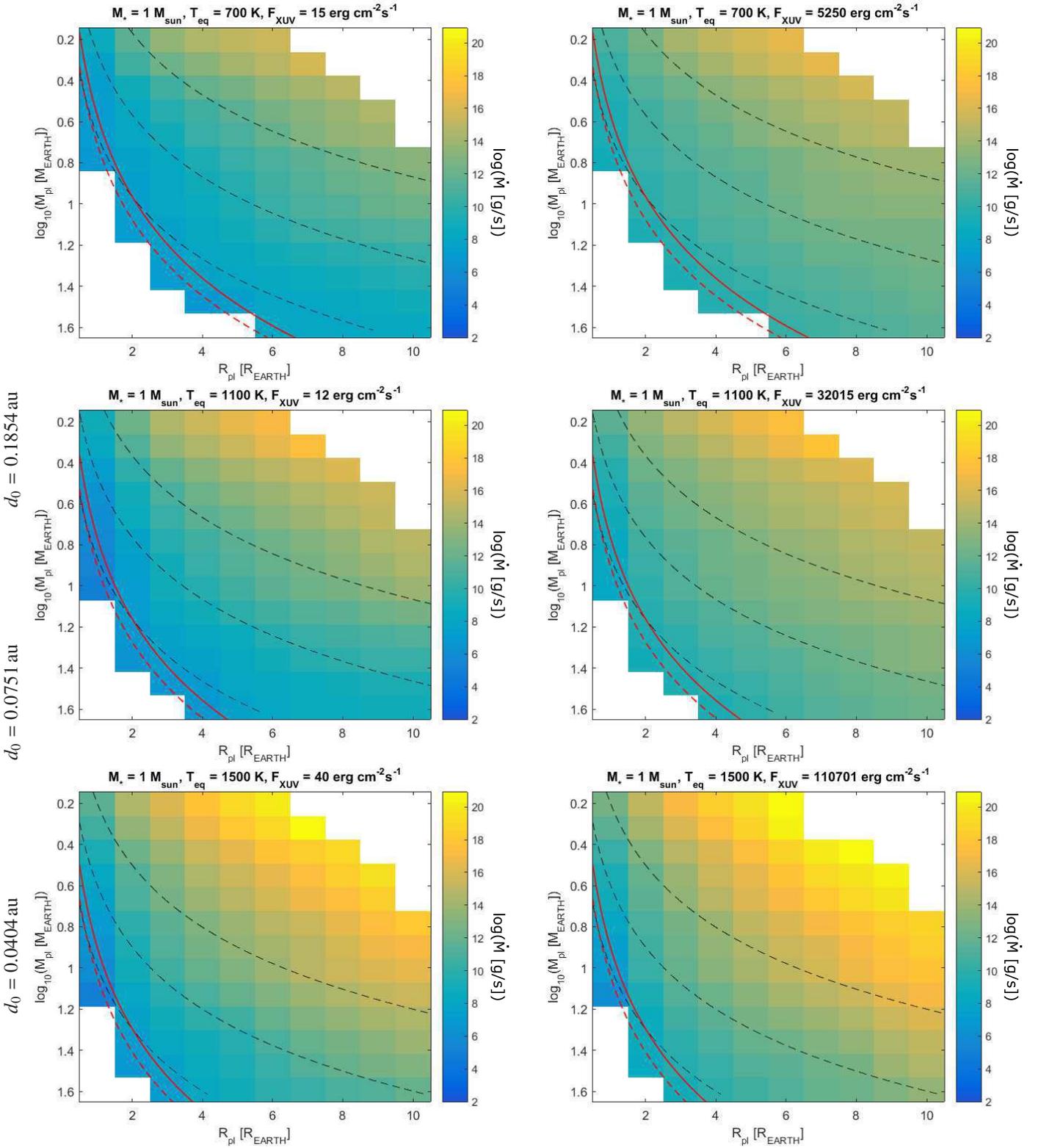}
\caption{Logarithm of the mass-loss rates (color coded) as a
function of planetary mass and radius. The adopted \Teq\ and
$F_{\rm XUV}$ values are given on the top of each panel. The
equilibrium temperature increases from top to bottom, while
$F_{\rm XUV}$ increases from left to right. All planets orbit a
1.0\,\Mo\ star. For reference, the dashed lines mark constant
$\Lambda$ values of 8, 20, and 50 (from top to bottom). {The red
lines indicate planets for which the Knudsen number at the upper
boundary is equal to 1 (solid line) and 10 (dashed line).}}
\label{fig:escapes}
\end{figure*}

Because of its relevance, e.g., in understanding planetary evolution, the mass-loss rate is one of the key output parameters of the modelling. Figure~\ref{fig:escapes} shows a few examples of how the mass-loss rates depend on planetary mass and radius for different \Teq\ and $F_{\rm XUV}$ values. All planets shown in Figure~\ref{fig:escapes} orbit a 1.0\,\Mo\ star. Appendix~\ref{apx:C} presents similar plots, but for planets orbiting stars more/less massive than 1.0\,\Mo. As expected, the highest mass-loss rates are found for the lowest gravity planets, whose atmospheres are in boil-off. With increasing $\Lambda$ (thus gravity), the mass-loss rates decrease first steeply and then more gradually. The dependence of the mass-loss rates on the stellar XUV flux tends to strengthen with increasing planetary mass.

{We further checked \emph{a posteriori} the validity of the
hydrodynamic equations to the modelled planets, namely whether the
atmosphere remains collisional (i.e., with efficient energy
redistribution) up to the sonic point. This condition is satisfied
if the Knudsen number $Kn$\,=\,$\lambda/l$\,$<$\,1, where
$\lambda$ is the mean free path of the gas and
$l$\,=\,$[\partial(\log{P})/\partial(r)]^{-1}$ is the
characteristic length scale. In Figure~\ref{fig:escapes} and in
the Figures in Appendix~\ref{apx:C}, we show lines corresponding
to $Kn$\,=\,1 and $Kn$\,=\,10. This indicates that the results we
obtained for the highest-density planets in our grid should be
taken with caution. However, these are anyway planets for which
the high bulk density disfavours the presence of a
hydrogen-dominated atmosphere.}

\begin{figure}[h!]
\includegraphics[width=\hsize]{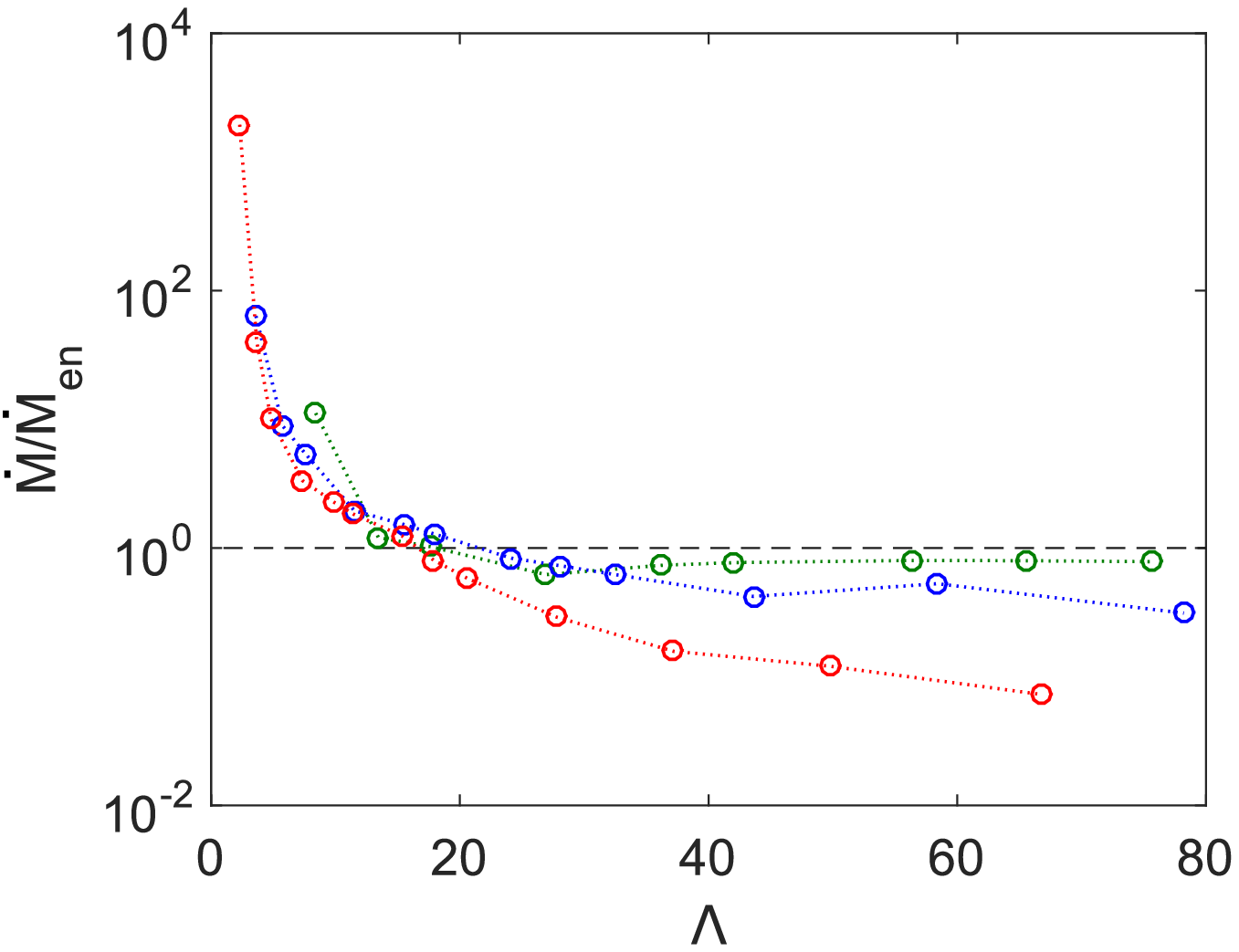}\\
\includegraphics[width=\hsize]{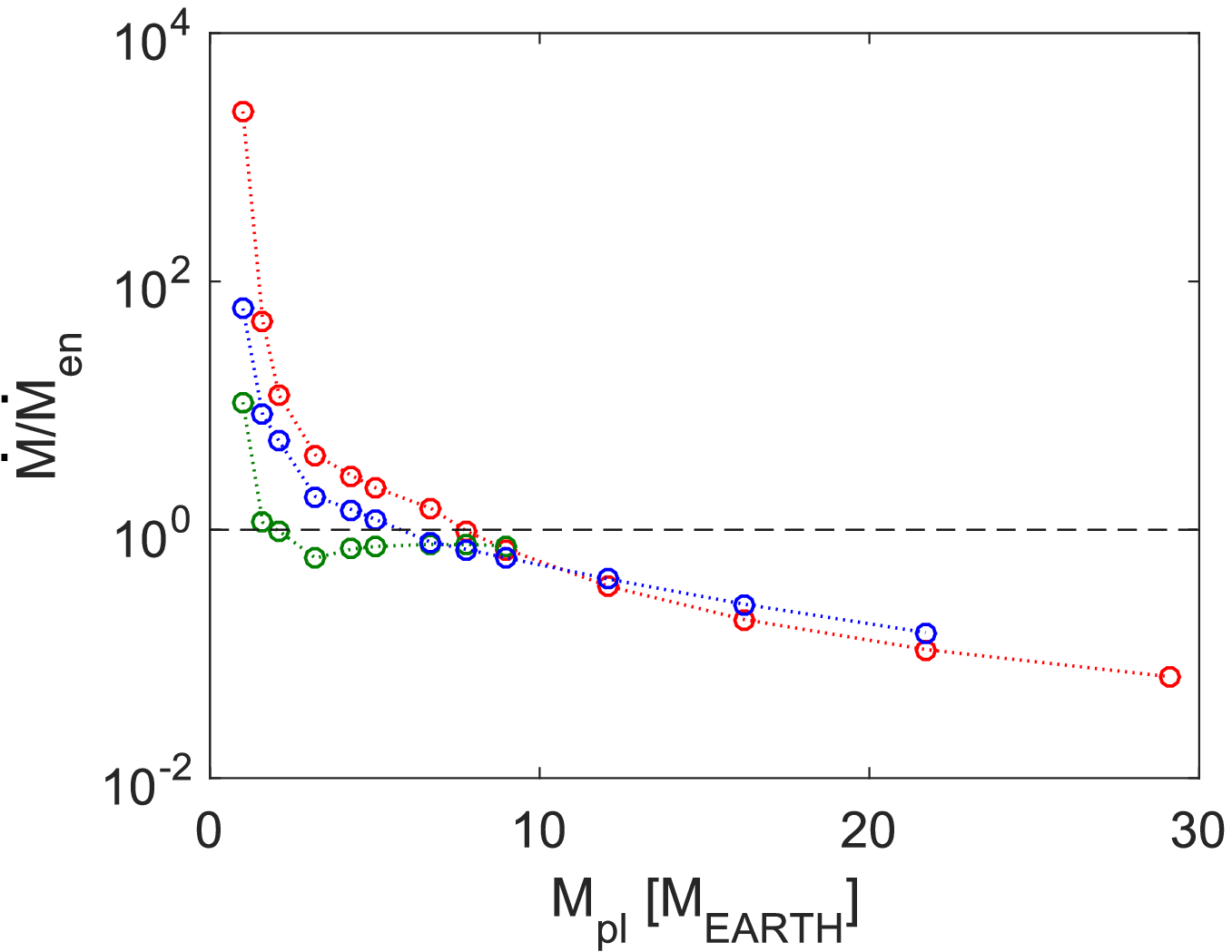}
\caption{Ratio between the mass-loss rates computed with the hydrodynamic model and the energy-limited formula as a function of $\Lambda$ (top panel) and $M_{\rm pl}$ (bottom panel). The green lines/circles are for systems with the following characteristics: $M_*$\,=\,1.0\,$M_{\odot}$, $T_{\rm eq}$\,=\,300\,K, $F_{\rm XUV}$\,=\,159.4\,\ergscm. The blue lines/circles are for systems with the following characteristics: $M_*$\,=\,1.0\,$M_{\odot}$, $T_{\rm eq}$\,=\,700\,K, $F_{\rm XUV}$\,=\,4900\,\ergscm. The red lines/circles are for systems with the following characteristics: $M_*$\,=\,1.0\,$M_{\odot}$, $T_{\rm eq}$\,=\,1100\,K, $F_{\rm XUV}$\,=\,30784\,\ergscm. The horizontal dotted line indicates the equality between the two values.}
\label{fig:lhy2len}
\end{figure}

In Figure~\ref{fig:lhy2len}, we compare the mass-loss rates as a function of $\Lambda$ (top) and planetary mass (bottom) obtained from the hydrodynamic modelling with those derived from the energy-limited formula
\begin{equation}
\dot{M}_{\rm en}=\frac{\pi\eta R_{\rm pl}R_{\rm eff}^2F_{\rm XUV}}{GM_{\rm pl}K}\,,
\label{eq:energyLimited}
\end{equation}
where the factor $K$ accounts for Roche-lobe effects \citep{erkaev2007}. By design, the energy-limited approximation works best for planets for which the atmosphere is hydrodynamic and the escape is driven by absorption of the stellar XUV flux, i.e., in blow-off. This implies that Equation~(\ref{eq:energyLimited}) overestimates the mass-loss rates for planets with hydrostatic atmospheres \citep[see e.g.,][]{fossati2018}. The top panel of Figure~\ref{fig:lhy2len} shows that, being in boil-off, the mass-loss rates for the lower-gravity planets are much higher than those predicted by Equation~(\ref{eq:energyLimited}). We also find that the $\dot{M}$/$\dot{M}_{\rm en}$ ratio decreases steeply with increasing $\Lambda$, having all other parameters constant. The value of $\Lambda$ at which the mass-loss rate computed with the hydrodynamic code becomes comparable to $\dot{M}_{\rm en}$ is about 20, in agreement with \citet{owen2016b} and \citet{fossati2017}. Figure~\ref{fig:lhy2len} shows also that Equation~(\ref{eq:energyLimited}) overestimates the mass-loss rates for planets with large $\Lambda$ values.

However, the plot in the top panel of Figure~\ref{fig:lhy2len}
might appear to be counterintuitive. This is because at large
$\Lambda$ values the hotter planets, thus more likely to have a
stable atmosphere, present mass-loss rates differing more from the
energy-limited approximation than the cooler ones. {This can be
explained by the fact that for a given value of $\Lambda$ hotter
planets have higher masses in comparison to cooler planets (see
Equation~(\ref{eq:lambda})), therefore in this plot the hotter
planets are more likely to have an hydrostatic atmosphere.} This
is clarified by the bottom panel of Figure~\ref{fig:lhy2len},
which shows that for the higher mass planets the difference
between the mass-loss rates computed by the hydrodynamic model and
with Equation~(\ref{eq:energyLimited}) is independent of \Teq.

Equation~(\ref{eq:energyLimited}) assumes that the entire stellar XUV energy input goes into driving the escape, but in reality part of this energy goes into running the chemical reactions, mainly ionisation of atomic hydrogen and dissociation of molecular hydrogen. Also, \citet{erkaev2015} showed that the
energy-limited formula neglects kinetic and thermal energy terms in the denominator, which can also be important for some star/planet parameters; i.e., even without including detailed chemistry or ionisation the results of Equation~(\ref{eq:energyLimited}) should be higher than the hydrodynamic mass-loss rates for XUV-driven outflows. This explains why for most planets the energy-limited approximation overestimates the mass-loss rates. However, this is not the case when there is a significant component of thermal escape, in which
case mass loss is driven partially by the stellar XUV flux and partially by the intrinsic planetary thermal energy. In this case, the mass-loss rates can significantly exceed those predicted by the energy-limited formula.

\subsection{Grid interpolation}\label{sec:interpolation}
%
{We developed a routine which interpolates the model results over
the grid parameter space considering planetary mass, planetary
radius, planetary equilibrium temperature, stellar mass, and
stellar XUV flux. For any system with parameters covered by the
grid, the routine extracts the density and outflow velocity, the
mass-loss rate, the value of maximum temperature, the effective
radius of XUV absorption, and the position of maximum dissociation
and ionisation.}

{The routine performs the interpolation in the following
consecutive steps.}
\begin{enumerate}
\item {For planets with input parameters [$\tilde{M_*}$,
$\tilde{T_{\rm eq}}$, $\tilde{F_{\rm XUV}}$, $\tilde{R_{\rm pl}}$,
$\tilde{M_{\rm pl}}$] the routine finds in the grid the two
closest values of stellar mass and equilibrium temperature
[$M_{*1}$, $M_{*2}$], [$T_{\rm eq1}$, $T_{\rm eq2}$].}

\item {For each of the four combinations of [$M_{*_i}$, $T_{\rm
eq_{\it j}}$], with $i,j$\,=\,1,2, the routine finds the two
closest values of the stellar XUV flux at the planetary orbital
separation $F_{\rm XUV_{\it ij}}^{k}$, with $k$\,=\,1,2 (i.e.,
eight $F_{\rm XUV}$ values).}

\item {For each set of [$M_{*_i}$, $T_{\rm eq_{\it j}}$, $F_{\rm
XUV_{\it ij}}^{k}$] the atmospheric parameters depend therefore
only on planetary radius and mass. Planets of the same mass have
different atmospheric properties for different equilibrium
temperatures, but we find similar atmospheric properties for
planets with similar $\Lambda$ values. Therefore, we substitute
planetary mass with $\Lambda$. At this point, the routine
interpolates the output parameters simultaneously over the pair
[\Rpl, $\Lambda$] for each of the eight sets of [$M_{*_i}$,
$T_{\rm eq_{\it j}}$, $F_{\rm XUV_{\it ij}}^{k}$]. However, for
planets beyond 0.1\,AU, the simultaneous interpolation on \Rpl\
and $\Lambda$ is not necessary, thus we reduced it to an
interpolation on $\Lambda$ only.}

\item {The routine interpolates the output parameters over $F_{\rm
XUV_{\it ij}}^{k}$.}

\item {The same equilibrium temperature for different stellar
masses occurs at rather different orbital separations (e.g.,
within the grid, a \Teq\ of $300$\,K corresponds to distances
ranging from 5$\times$10$^{-3}$ to 1.52\,AU). Our analysis
\citep[see Section~\ref{sec:discussion} and
also][]{kubyshkina2018} indicates that between \Teq\ and $d_0$,
the latter has the larger influence on the results, thus for the
interpolation we substitute \Teq\ with $d_0$. Therefore, the
routine interpolates the output parameters over $d_{0_j}$ for the
pair of $M_{*_i}$.}

\item {The routine interpolates the output parameters over
$M_{*_i}$.}
\end{enumerate}

{We developed this routine keeping in mind that the size of the
grid will increase in the future, therefore the need of an
interpolation routine capable of quickly handling the addition of
grid points. This is why we avoided to use complicated,
multi-dimensional interpolation functions that would require
recomputing every time a new model is added to the grid.}

{Because the output parameters behave differently as a function of
the input parameters, for almost each interpolation step and
almost each output parameter, we employ a different function. For
the mass-loss rates and the density at the Roche radius the
routine interpolates over \Rpl, $\Lambda$, and $F_{\rm XUV}$
according to}
\begin{equation}
\label{eq:approximation1}
\ln{X} = a + b\,\ln{R_{\rm pl}}\,,
\end{equation}
\begin{equation}
\label{eq:approximation2}
\ln{X} = c + d\,\ln{\Lambda}\,,
\end{equation}
and
\begin{equation}
\label{eq:approximation3}
\ln{X} = e + f\,\ln{F_{\rm XUV}}\,,
\end{equation}
{where X is either the mass-loss rate or the density at the Roche
radius and the coefficients depend on the other system parameters.
For planets in boil-off, Equations~(\ref{eq:approximation1}) and
(\ref{eq:approximation2}) are not accurate enough, therefore we
use a piece-wise polynomial interpolation with input and output
parameters in logarithmic scale. The interpolation of the
mass-loss rates and the density at the Roche radius over the other
input parameters (i.e., $d_0$ and $M_*$) is done on the basis of a
linear function.}

{For the outflow velocity, we perform the interpolation on the
[\Rpl, $\Lambda$] pair using a third order polynomial function,
which becomes linear for $\Lambda$\,$\gtrsim$\,20. For the
interpolation of the outflow velocity over the other input
parameters we use a linear function.}

{For the interpolation of the maximum temperature over each input
parameter, we employ a linear function. The routine interpolates
the position of the maximum dissociation and ionisation and of the
\Reff\ value on the [\Rpl, $\Lambda$] pair using a linear function
for $\Lambda$\,$\gtrsim$\,20, while for smaller $\Lambda$ values
we interpolate over $\Lambda$ using a function of the form $a / (b
+ \Lambda)$, where the coefficients $a$ and $b$ depend on the
other input parameters. For the interpolation of the maximum
temperature, maximum dissociation and ionisation, and \Reff\ value
over the other input parameters (i.e., $F_{\rm XUV}$, $d_0$, and
$M_*$) we employ a linear function.}

\begin{figure}
\includegraphics[width=\hsize]{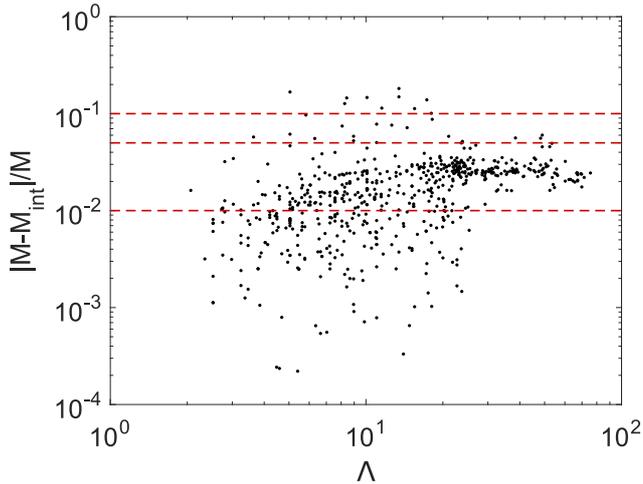}
\caption{{Relative deviation of the interpolated mass-loss rates
($\dot{M}_{\rm int}$) from the computed ones ($\dot{M}$) as a
function of $\Lambda$. The horizontal red lines indicate
deviations of 1\%, 5\%, and 10\%.}} \label{fig:ierr}
\end{figure}

{We performed two tests to validate the interpolation routine. We
first compared the results obtained from the models with those
derived using the interpolation for 500 systems randomly
distributed across the grid. We found an agreement of better than
5\% in 95\% of the cases, while for the remaining systems the
agreement was better than 20\% (Figure~\ref{fig:ierr}).}

{The second test is dedicated to check the validity of
substituting \Teq\ with $d_0$ for the interpolation. We used real
planets for this test, namely Kepler-11\,b, GJ\,436\,b,
HAT-P\,26\,b, HD97658\,b, GJ\,3470\,b, HAT-P\,11\,b, and
55\,Cnc\,e, which lie within our grid boundaries, and compared the
mass-loss rates obtained with direct modelling and interpolation.
Figure~\ref{fig:lhy-lambda} shows the results of this comparison,
indicating that we obtain an excellent agreement for all of them.}
This validates our choice of interpolating on the orbital
separation rather than on the planetary equilibrium temperature.
We run the same test, but this time interpolating on the planetary
equilibrium temperature, instead of orbital separation, obtaining
significantly larger discrepancies.
\section{Discussion}\label{sec:discussion}
We discuss here in more detail how {the results of the grid}
depend on the input parameters (Section~\ref{sec:output-input})
and briefly explore one of its possible future applications
(Section~\ref{sec:evolution}). The reader not interested in the
technicalities of the results can skip to
Section~\ref{sec:evolution}, which is independent from what is
described in Section~\ref{sec:output-input}.
\subsection{Behaviour of the atmospheric parameters as a function of input parameters}\label{sec:output-input}
The grid allows for the detailed description of how the atmospheric structure changes with respect to the input system parameters. The behaviour of the main output parameters can be separated into common patterns. We found a common behaviour between {\it i}) mass-loss rates and densities of the atmospheric species (Section~\ref{sec:massloss}); {\it ii}) effective radius of the stellar XUV absorption and position of the maximum dissociation and ionisation (Section~\ref{sec:barriers}); {\it iii}) outflow velocity and atmospheric temperature (Section~\ref{sec:velocity}).
\subsubsection{Mass-loss rates and densities of the atmospheric species}\label{sec:massloss}
\begin{figure*}
\includegraphics[width=\hsize,clip]{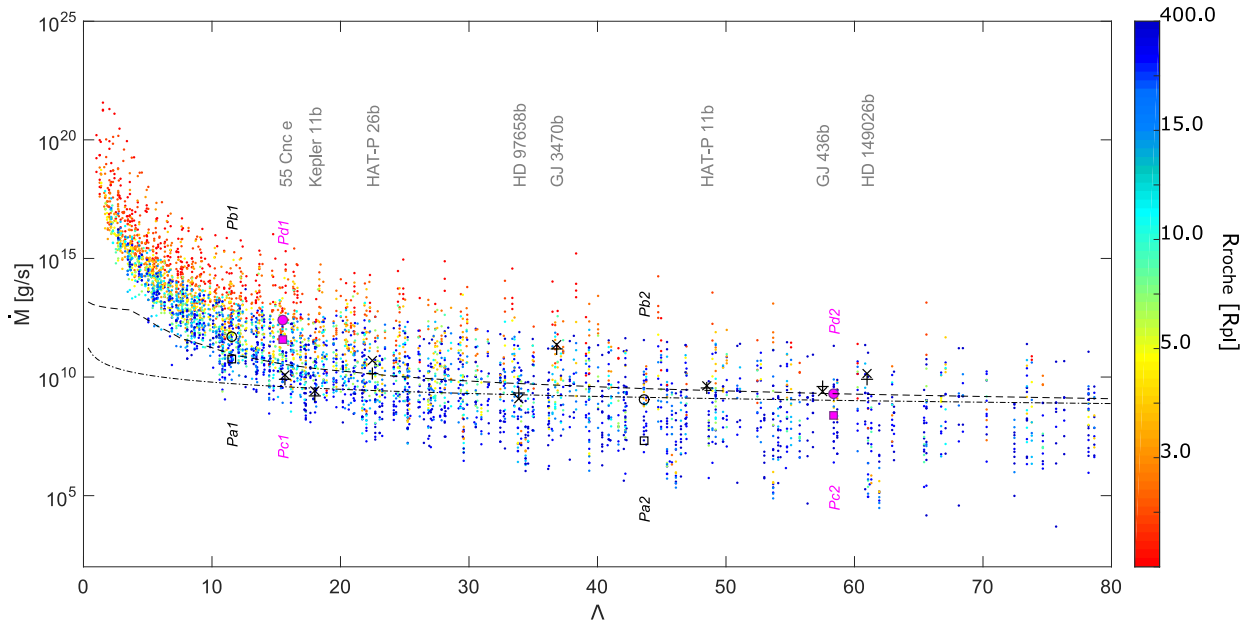}
\caption{Atmospheric mass-loss rate $\dot{M}$ as a function of $\Lambda$ for all computed planets. The color code indicates the planetary Roche radius in units of \Rpl. The position of the test planets listed in Table~\ref{tab:testP} is shown by black squares (Pa1 and Pa2), black circles (Pb1 and Pb2), purple squares (Pc1 and Pc2), and purple circles (Pd1 and Pd2). The lines indicate the predictions obtained by using the energy-limited formula for the Pb test planets varying planetary mass only and assuming the value of \Reff\ equal to the planetary radius (dash-dotted line) or equal to the value derived from the grid (dashed line). The black crosses and plus signs mark the escape rates estimated for some of the known transiting exoplanets using the interpolation routine and direct hydrodynamic calculations, respectively.}
\label{fig:lhy-lambda}
\end{figure*}

One of the major parameters controlling the long-term evolution of a planetary atmosphere is the mass-loss rate $\dot{M}$, which strongly depends on the planetary gravity and orbital separation, thus the equilibrium temperature. Figure~\ref{fig:lhy-lambda} shows the dependence of $\dot{M}$ on $\Lambda$. Within the parameters covered by our grid, the mass-loss rate varies by several orders of magnitude, with planetary gravity and Roche lobe radius being among the main parameters controlling it.

For the planets with $\Lambda$ values smaller than about 20, the
escape rates reach extreme values of up to 10$^{20}$\,\gs, due to
a combination of low planetary gravities and high equilibrium
temperatures (i.e., boil-off). The atmospheres of these planets
are therefore characterised by strong thermal escape and
inefficient XUV heating. The escape rates for the majority of
these planets lie above the predictions of the energy-limited
formula, and the energy {budget of the atmosphere is} dominated by
adiabatic cooling. In first approximation, for a given stellar
mass, \Teq, and stellar XUV flux, the dependence of the mass-loss
rate on $\Lambda$ can be described by
Equation~(\ref{eq:approximation2}). Such high escape rates would
imply a rapid escape of the atmosphere (we provide a practical
example of this in Section~\ref{sec:evolution}).

With increasing $\Lambda$, the efficiency of XUV heating increases with XUV penetration depth (see Figure~\ref{fig:profiles}) and the mass-loss rates become strongly dependent on the stellar XUV flux. This further dependence of the mass-loss rates is partly responsible for the increased spread in mass-loss rates at large $\Lambda$ values. We also note that the spread increases with decreasing stellar mass, due to the decreasing orbital separation for the same temperature. The dependence of the mass-loss rates on the stellar XUV flux can be roughly described by a linear function, in agreement with the energy-limited formula.

Figure~\ref{fig:lhy-lambda} shows also that the Roche radius plays
an important role almost exclusively when it lies below about
15\,\Rpl\  (in Figure~\ref{fig:lhy-lambda}, the dark blue color
corresponds to Roche radii ranging from 15 up to 400\,\Rpl). It is
important to remind that the Roche radius is tightly related with
the orbital separation and the smallest Roche radii can be reached
just for the shortest star-planet distances. As an example, within
our grid, the smallest Roche radii ($<$3\,\Rpl) are reached only
for planets lying less than 0.06\,AU from the host star, while
Roche radii of 15\,\Rpl\ are found for planets orbiting up to
0.3\,AU from the host star. The mass-loss rates increase with
decreasing Roche radii {because a smaller Roche radius moves the
sonic point closer to the planet, i.e., to regions of higher
density, which leads to an increase in mass-loss rate.} In
addition, the Roche radius decreases with decreasing orbital
separation (see Equation~(\ref{eqn:Rroche})), thus increasing
\Teq\ and XUV irradiation.

To better illustrate how the escape rates change with input parameters, we set eight test planets (called Pa1, Pa2, Pb1, Pb2, Pc1, Pc2, Pd1, Pd2), whose parameters are listed in Table~\ref{tab:testP}. The numbers ``1'' and ``2'' separate planets by mass, where the planets identified by the number ``1'' have the lower mass. The difference between the ``a'' \& ``b'' and ``c'' \& ``d'' planets is the stellar mass, which is higher for the former, while the difference between the ``a'' \& ``c'' and ``b'' \& ``d'' planets is the stellar XUV flux, which is higher for the latter. To have just the planetary mass controlling the value of $\Lambda$, the eight planets have the same equilibrium temperature (700\,K) and planetary radius (3\,\Re). Figure~\ref{fig:lhy-lambda} indicates the position of the eight planets in the $\dot{M}$ vs $\Lambda$ plane.
\begin{table*}
\caption{Test planets considered for the discussion of the results. The Roche radius here is defined from the center of the planet.}
\label{tab:testP}
\centering
\begin{tabular}{c|c|c|c|c|c|c|c|c|c|c|c}
  \hline
  \hline
  ID & $M_*$ & \Teq & $F_{\rm xuv}$ & \Rpl & \Mpl & $\Lambda$ & \roche & $R_{\rm eff}$ & $\dot{M}$ & $T_{\rm max}$ & $V_{\rm max}$ \\
     & [\Mo] & [K]  & [\ergscm]     & [\Re]& [\Me]&           & [\Rpl] & [\Rpl]        & [\gs]     & [K]           & [\kms]        \\
  \hline
  Pa1 & 1.3 & 700 & 19.5    & 3 & 3.2  & 11.5 & 29.5 & 6.76 & $5.8\times10^{10}$ & 700  & 0.48 \\
  Pa2 & 1.3 & 700 & 19.5    & 3 & 12.1 & 43.6 & 45.9 & 1.90 & $2.1\times10^7$    & 1801 & 1.07 \\
  Pb1 & 1.3 & 700 & 1110.5  & 3 & 3.2  & 11.5 & 29.5 & 4.09 & $4.9\times10^{11}$ & 700  & 0.89 \\
  Pb2 & 1.3 & 700 & 1110.5   & 3 & 12.1 & 43.6 & 45.9 & 1.53 & $5.3\times10^9$    & 2318 & 2.25 \\
  Pc1 & 0.4 & 700 & 18.5    & 3 & 4.3  & 15.5 & 2.9  & 1.43 & $3.7\times10^{11}$ & 700  & 0.39 \\
  Pc2 & 0.4 & 700 & 18.5    & 3 & 16.2 & 58.4 & 4.6  & 1.15 & $2.4\times10^8$    & 3331 & 0.51 \\
  Pd1 & 0.4 & 700 & 16731 & 3 & 4.3  & 15.5 & 2.9  & 1.04 & $2.5\times10^{12}$ & 737  & 0.76 \\
  Pd2 & 0.4 & 700 & 16731 & 3 & 16.2 & 58.4 & 4.6  & 1.01 & $2.0\times10^{9}$ & 4370 & 1.37 \\
  \hline
\end{tabular}
\end{table*}

As expected, an increase in the XUV stellar flux (i.e., Pa$\rightarrow$Pb or Pc$\rightarrow$Pd) or decrease in $\Lambda$ (i.e., Px2$\rightarrow$Px1, where x is any of a, b, c, or d) leads to an increase in the mass-loss rates. Figure~\ref{fig:lhy-lambda} shows also that a decrease in stellar mass (i.e., Pa$\rightarrow$Pc or Pb$\rightarrow$Pd) leads as well to an increase in mass-loss rates. This is because, to maintain the same \Teq, planets orbiting around the lower mass star lie at a closer distance, thus have a smaller Roche radius and for the reasons described above have a higher mass-loss rate.

For the Pb planets, Figure~\ref{fig:lhy-lambda} also compares the mass-loss rates with those predicted by the energy-limited formula assuming two different values for the effective radius and a heating efficiency of 15\%, as for the hydrodynamic calculations. Since with decreasing $\Lambda$ the effective radius moves farther away from the planet, the distance between the two lines in Figure~\ref{fig:lhy-lambda} increases with decreasing $\Lambda$. The dashed line in Figure~\ref{fig:lhy-lambda} presents a clear bend at $\Lambda$\,$\approx$\,4, which is caused by the fact that at small $\Lambda$ values the effective radius reaches the Roche radius. Figure~\ref{fig:lhy-lambda} indicates that at low $\Lambda$ values the energy-limited approximation significantly underestimates the escape rates (2--3 orders of magnitude; comparison to the black circles), while at large $\Lambda$ values the approximation overestimates the escape rates by about one order of magnitude (see also Figure~\ref{fig:lhy2len}).

The atmospheric densities (at the Roche radius) behave similarly to the mass-loss rates. The only small difference is found for planets with large $\Lambda$ values (densities decrease steeper). This is because the escape rates are calculated from the product of the atmospheric density and outflow velocity at the Roche radius, where the velocity increases with increasing Roche radius, which increases with increasing planetary mass.

The dependence of the escape rates and atmospheric densities on
planetary mass is essentially the same as that on $\Lambda$,
though slightly less pronounced. The dependence of these
parameters on \Teq\ is similar to what is displayed by the color
code in Figure~\ref{fig:lhy-lambda} and it can be described by
using a log-linear approximation of the form $\log{X} = c_1 +
c_2\,T_{\rm eq}$, where X is either the mass-loss rate or the
atmospheric density, and $c_1$ and $c_2$ are coefficients, which
depend on the system parameters. {This follows the high $\Lambda$
limit of the Parker wind problem.}
\subsubsection{Effective radius and position of the maximum dissociation and ionisation}\label{sec:barriers}
\begin{figure}[h!]
\includegraphics[width=\hsize,clip]{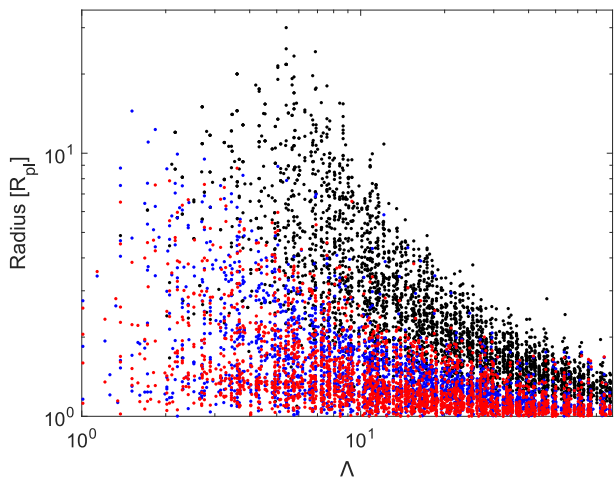}
\caption{{Effective radius (black), position of the maximum
dissociation (blue), and position of the maximum ionisation (red)
as a function of $\Lambda$, for all planets in the grid.}}
\label{fig:reff-dis-ion}
\end{figure}

We discuss here the behavior of three closely related parameters: the effective radius (Equation~(\ref{eqn:reff})) and the position of the maximum dissociation and ionisation. As we defined in Section~\ref{sec:results}, the position of the maximum dissociation and ionisation corresponds to the position of the maximum of $\nh$ and $\nhp$, respectively. Figure~\ref{fig:reff-dis-ion} shows the position of these three quantities as a function of $\Lambda$. For most planets, they lie close to each other, except for planets with small $\Lambda$ values (i.e., $\lesssim$20), where the effective radius significantly exceeds the other two. At small $\Lambda$ values the effective radius can be up to ten times larger than the position of the maximum dissociation/ionisation, which stays close to one another for the whole interval of parameters. For $\Lambda$ values above $\sim$50, the difference between the position of maximum dissociation/ionisation and the effective radii lies roughly within 30\%. We also found that for these planets the difference between the three values decreases slightly with increasing stellar XUV flux and planetary mass, where the latter dependence is caused by the gradual compression of the atmosphere with increasing planetary mass.

The possible range of values found in the grid for the effective radius and the position of the maximum dissociation and ionisation increases significantly with decreasing $\Lambda$ and it reaches the maximum close to $\Lambda$\,=\,5. Here the effective radius reaches the Roche radius, which decreases with decreasing $\Lambda$. The position of maximum dissociation/ionisation reaches this artificial border at smaller $\Lambda$ values, i.e., $\approx$2.

\begin{figure}[h!]
\includegraphics[width=\hsize,clip]{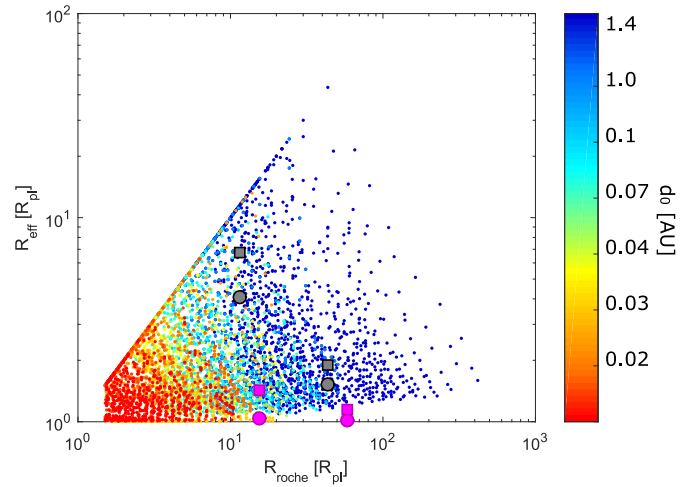}
\caption{{Effective radius \Reff\ as a function of \roche. Each
point is color-coded with respect to the orbital semi-major axis.
The black and purple circles and squares indicate the position of
the eight test planets following the same symbols as in
Figure~\ref{fig:lhy-lambda}. The circle and square purple symbols
largely overlap.}} \label{fig:rdis-lambda}
\end{figure}

We found also a clear dependence of the three quantities on orbital separation. To highlight this, Figure~\ref{fig:rdis-lambda} shows the effective radius as a function of the Roche radius and of the orbital separation. Again, for each given Roche radius, the effective radius presents an upper limit, which is clearly given by \roche\ itself. The spread in effective radii also increases with increasing \roche\ and $d_0$. This can be understood as follows. At short orbital separations, \roche\ is generally small and therefore there is only a small range of possible effective radii. At large orbital separations, instead, depending on the planetary and stellar masses, there is a much wider range of possible Roche radii within which the position of maximum dissociation can lie. However, Figure~\ref{fig:rdis-lambda} shows that in our grid there are very few planets with large Roche and effective radii, namely those with a rather low-density and orbiting far from the host stars. For these planets, \roche\ is large and the stellar XUV flux is too weak for the effective radius to be close to \Rpl.

Finally, there are a few remarks that need to be made regarding
these parameters. Unlike the definition of the effective radius
(given by Equation~(\ref{eqn:reff})), the definitions of the
position of maximum dissociation and ionisation are not univocal.
In addition, in some cases, the dissociation and ionisation
profiles are very smooth, which makes the position of maximum
dissociation and ionisation somewhat dependent on their
definitions. It is artificial to consider that the effective
radius and/or the position of the maximum ionisation and
dissociation are located at the Roche lobe if their position moves
beyond it. It is therefore important to keep in mind that for some
particular planets these {values are indicative,} rather than
sharp results, and that comparisons with the literature and/or
with future studies shall consider these issues.
\subsubsection{Outflow velocity at the Roche radius and maximum atmospheric temperature}\label{sec:velocity}
\begin{figure}
\includegraphics[width=\hsize,clip]{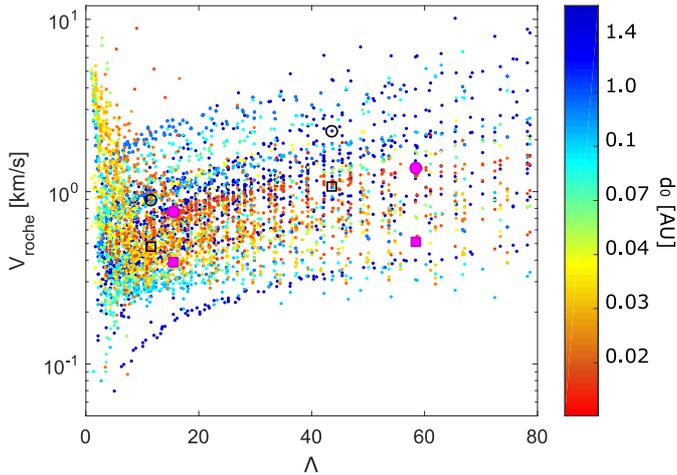}
\caption{{Outflow velocity at the Roche radius $V_{\rm roche}$ as
a function of $\Lambda$. Each point is color coded with the value
of the orbital semi-major axis. The black and purple circles and
squares indicate the position of the eight test planets following
the same symbols as in Figure~\ref{fig:lhy-lambda}.}}
\label{fig:vmax-lambda}
\end{figure}

We define the outflow velocity $V_{\rm roche}$ as the velocity of the atmospheric particles crossing the Roche lobe. Figure~\ref{fig:vmax-lambda} shows the velocity at the Roche radius as a function of $\Lambda$ and orbital semi-major axis. For planets with $\Lambda$ values $\gtrsim$10, the velocity of the escaping material grows linearly with increasing $\Lambda$. This is caused by the gradual increase of the Roche radius with $\Lambda$, which allows for longer acceleration distances and for a smaller planetary gravitational pull at the Roche radius. This lies partly at the origin of the connection between the velocity at the Roche radius and orbital semi-major axis.

For planets with very small $\Lambda$ values, the velocity behaves exactly in the opposite way, it increases with decreasing $\Lambda$ as a result of the low gravity and small Roche radius. We found that the planets for which $V_{\rm roche}$ is larger than 2\,\kms\ have a Roche radius located closer than 5\,\Rpl, while for more extreme planets with $V_{\rm roche}$ greater than 4\,\kms, the Roche radius is always smaller than 1.5\,\Rpl.

The dependence of the outflow velocity on {the} XUV flux is
similar to that of the mass-loss rates. To illustrate this, we
added the position of the eight test planets to
Figure~\ref{fig:vmax-lambda}. The influence of the equilibrium
temperature on the velocity at the Roche radius is however
unclear, possibly because variations in the equilibrium
temperature imply simultaneous changes in the atmospheric
structure and in the orbital separation, thus in the Roche radius.
In \citet{kubyshkina2018}, we showed that, keeping planetary mass
and radius fixed, the effects of \Teq\ variations on the velocity
are negligible for planets with $\Lambda$ greater than 20, while
for planets with lower $\Lambda$ the velocity may significantly
decrease with increasing temperature.

We found that the maximum value of the atmospheric temperature $T_{\rm max}$ behaves similarly to the outflow velocity, except for planets with small $\Lambda$ values ($\lesssim$15--20). For these planets, the atmospheric temperature profile is characterised by strong adiabatic cooling, which implies that the maximum temperature is equal to the equilibrium temperature. For planets with larger $\Lambda$ values, the maximum atmospheric temperature increases almost linearly with $\Lambda$, similarly to the velocity, but it further presents a more pronounced dependence on the stellar XUV flux.
\subsection{Atmospheric evolution of the high-density exoplanets CoRoT-7\,b and HD219134\,b,c}\label{sec:evolution}
We present here a direct application of the grid, namely a simple
scheme allowing to infer the evolution of a planetary atmosphere
subject to mass-loss, with the mass-loss rates extracted from the
grid. Thanks to the dense grid and possibility to interpolate
across it, one can {quickly} derive high-resolution evolutionary
tracks of planets with parameters contained in the grid. The
advantage is that the tracks are obtained making use of mass-loss
rates computed with an hydrodynamic code, rather than more
approximate methods, such as the energy-limited formula. As an
example, we study the past evolution of the possible primary
atmospheres of the close-in high-density planets CoRoT-7\,b
\citep{leger2009,valencia2010,leitzinger2011,mura2011,barros2014}
and HD219134\,b,c \citep{motalebi2015,vogt2015,gillon2017}.

%
\subsubsection{Planetary evolution modelling scheme}
{We first assume that the orbital separation and stellar mass do
not change with time. It follows that the mass-loss rates at every
moment in time depend only on the planetary radius and the amount
of bolometric and XUV stellar irradiation.}

{To infer planetary atmospheric mass fractions ($f_{\rm at}$), we
employ the model described by \citet{johnstone2015}, which relates
$f_{\rm at}$ with planetary mass and radius. However, the
approximate relation between these three quantities given by
\citet{johnstone2015} was obtained considering atmospheres having
a density of 5$\times$10$^{12}$\,cm$^{-3}$ and a temperature of
250\,K at the base of the simulation, which significantly differ
from the conditions of our planets. Therefore, we used the code
employed by \citet{johnstone2015} to directly compute a grid of
$f_{\rm at}$ values ((hereafter called $f_{\rm at}$-grid) for
planets with mass and radius in the range of interest of this work
and interpolate among the grid points. For each planet considered
in the $f_{\rm at}$-grid, the core radius $R_{\rm core}$ was
derived assuming an Earth-like density.}

{One of the key parameters to set to simulate the atmospheric
evolution is the initial planetary radius. Various accretion
models provide an estimate of the initial planetary radius, thus
atmospheric mass accreted by the planet while embedded in the
protoplanetary nebula \citep[e.g.,][]{stokl2016}, but the results
are rather model dependent and small variations may affect the
tracks. We approach this problem in a more empirical way: we
calculate tracks assuming three different initial radii and see
{\it a posteriori} which is the impact of the assumption of the
initial radius on the evolutionary tracks. As initial radii for
each planet, we assume the values obtained by setting $\Lambda$
equal to 3, 5, and 10.}

{For late-type stars, the stellar XUV flux depends on the stellar
mass and rotation period, where the latter is time-dependent. In
this work, we assume that the rotation period varies with time as
\citep{mamajek2008}}
\begin{equation}
\label{eq:Trot}
P_{\rm rot}=0.407\,[(B-V)_0-0.495]^{0.325}\,\tau^{0.566}\,,
\end{equation}
where $P_{\rm rot}$ is the rotation period (in days), $\tau$ is
the stellar age (in Myr), and $(B-V)_0$ is the reddening-free
stellar color. {Equation~(\ref{eq:Trot}) represents the average
approximation based on a large set of late-type dwarfs, but in
reality the rotation tracks of stars are non-unique, which leads
to different evolutionary tracks of the XUV radiation
\citep{johnstone2015rot,tu2015}. In the most general case, this
results in a variety of planetary atmosphere evolution tracks, but
for the present analysis that is limited to small, close-in
planets we restrict ourselves to the approximation given by
Equation~\ref{eq:Trot}.}

{We set the relevant stellar parameters as follows. The stellar
X-ray flux can then be inferred from the rotation period as
\citep{wright2011}
\begin{equation}
\label{eq:Xluminosity}
\frac{L_{\rm X}}{L_{\rm bol}} =
    \begin{cases}
    CR_{\rm 0,sat}^{\beta} & \text{if $R_0\leq R_{\rm 0,sat}$}\\
    CR_0^{\beta}  & \text{if $R_0>R_{\rm 0,sat}$}\,,
    \end{cases}
\end{equation}
where $L_{\rm X}$ is the X-ray stellar luminosity, $L_{\rm bol}$ is the bolometric luminosity, $C$\,=\,8.68$\times$10$^{-6}$ and $\beta$\,=\,$-$2.18 are empirical constants \citep{pizzolato2003,wright2011}, and $R_{\rm 0,sat}$\,=\,0.13 is the saturation threshold corresponding to the Rossby number $R_0$. This last quantity is the ratio between the stellar rotation period and the convective turnover time \citep[$T_{\rm conv}$;][]{wright2011}
\begin{equation}
\label{eq:Tconv}
\log{T_{\rm conv}}=1.16-1.49\,\log{M_*}+0.54\,\log^2{M_*}\,.
\end{equation}
The EUV stellar luminosity is then derived from the X-ray luminosity using Equation~(\ref{eq:X2EUV}). To account for how the bolometric luminosity and equilibrium temperature change with time, we employ the MESA Isochrones and Stellar Tracks \citep[MIST;][]{paxton2018}.}

We begin the evolution at the age of 5\,Myr, which is
approximately the typical lifetime of protoplanetary disks
\citep{mamajek2009}. However, since we are in general interested
in Gyrs-old planets, the exact initial time for the evolution has
no significant effect on the results. {Having set the initial
planetary radius, $f_{\rm at}$, and XUV stellar flux at the
planetary orbital separation, we extract the mass-loss rate from
our grid, which we then use to derive how much mass is lost during
the first time step. At this point, we derive the new planetary
atmospheric mass that we convert into a planetary radius by
interpolating over the $f_{\rm at}$-grid, and begin the cycle
again, updating at each time step the stellar XUV flux. We finally
obtain atmospheric evolutionary tracks by choosing small-enough
time steps, which in our case adapt to the mass-loss rates by
ensuring that the maximum mass loss within one time step is
smaller than 1\% of the planetary mass, and by repeating this
procedure up to the desired age (e.g., age of the given system) or
till the planetary radius has reached the core radius. We ignore
gravitational contraction and radioactive decay that contribute to
increase the equilibrium temperature during the first phases of
evolution.}
\subsubsection{Results for CoRoT-7\,b}
CoRoT-7\,b has a mass of 5.74\,\Me\ and a radius of 1.585\,\Re,
which indicates a rocky composition and a lack of a
hydrogen-dominated envelope \citep{leger2009,mura2011,barros2014}.
The planet orbits an {active} early K-type star
($M_*$\,=\,0.93\,\Mo, $R_*$\,=\,0.87\,\Ro, \Teff\,=\,5275\,K) at a
distance of 0.0172\,AU, which corresponds to a period of
0.853\,days. The planet has an equilibrium temperature of 1756\,K
and the age of the system has been estimated to be
1.5$\pm$0.3\,Gyr.

\begin{figure}
\includegraphics[width=\hsize,clip]{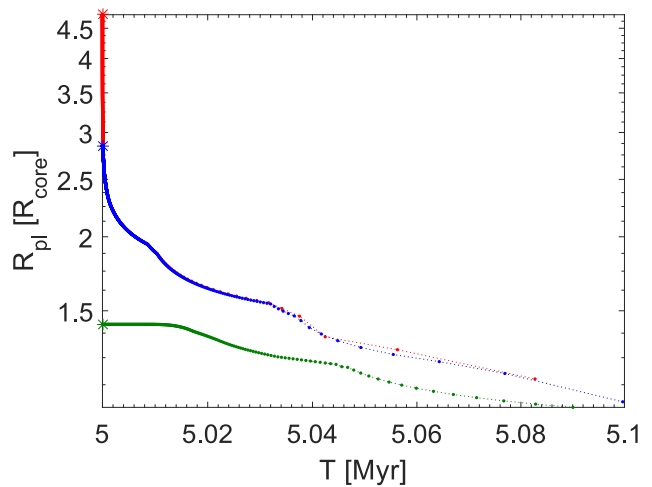}
\caption{{Evolution of the planetary radius of CoRoT-7\,b as a
function of time. The colors indicate different initial radii,
marked by the asterisks, which correspond to the values obtained
by setting $\Lambda$\,=\,3 (red), 5 (blue), and 10 (green). The
small dots placed along each line indicate the time steps.}}
\label{fig:corot7b}
\end{figure}

Figure~\ref{fig:corot7b} shows the evolution of the planetary
radius as a function of time obtained for each of the three tested
initial conditions corresponding to planetary radii of 8.25\,\Re\
($f_{\rm at} \approx 6\times 10^{-2}$), 4.95\,\Re\  ($f_{\rm at}
\approx 2\times 10^{-2}$), and 2.47\,\Re\  ($f_{\rm at} \approx
6\times 10^{-4}$). The two tracks starting with the largest
planetary radius converge {quickly} to the same point, implying
that the result is independent of the initial condition. This is
because the first part of the evolution is dominated by boil-off.
The track starting with the smaller planetary radius instead does
not present clear signs of a boil-off phase, but still {leads} to
a rapid complete escape of the atmosphere. The tracks indicate
that the planet is supposed to have {completely lost its
hydrogen-dominated envelope within an extremely short time of
about 0.1\,Myr.}

The most significant changes in the size of the atmosphere occur during the first $10^{-2} - 10^{-1}$\,Myr, when the atmosphere lies in the boil-off regime, which is consistent with what was found by \citet{owen2016b}. Once the radius has reached about 2\,\Re, the escape is driven by the stellar XUV flux, which for CoRoT-7\,b is very intense since the system is still rather young and the planet has a very short orbital period. The complete escape of the atmosphere is so fast that the initial stellar rotation rate \citep[see, e.g.,][]{tu2015} does not play a significant role. We therefore estimate that CoRoT-7\,b has lost its primary atmosphere, assuming it had accreted one to begin with, within a maximum time of about 0.1\,Myr.

Atmospheric escape for CoRoT-7\,b has been previously studied by \citet{jackson2010} and \citet{leitzinger2011}. Both inferred the planet's mass-loss rate over time using the energy-limited approximation, accounting for the Roche lobe effect \citep{erkaev2007}, which greatly underestimates the mass-loss rates at the beginning of the planet's evolution.

\citet{jackson2010} considered the effects of the possible planetary migration through orbital tidal decay, a wide range of initial radii (up to a gas giant), an effective radius of 3\,$R_{\rm pl}$, various heating efficiencies up to 100\%, and applied the scaling laws for the stellar XUV flux of \citet{ribas2005}. They arrived at the conclusion that CoRoT-7\,b could have started its evolution as a gas giant, with a mass of up to 200\,\Me. In case, instead, CoRoT-7\,b has always been a rocky planet, they suggested that it could have lost up to half of its mass through surface melting, outgassing, and subsequent escape of the secondary atmosphere.

\citet{leitzinger2011} did not account for planetary migration, but employed more realistic heating efficiencies of 10-25\% and stellar irradiation levels consistent with those adopted in our work. Their calculations led them to exclude that CoRoT-7\,b had started its evolution as a gas giant planet, with a mass similar or larger than that of Saturn, otherwise the planet would still host a significant hydrogen-dominated envelope, which is excluded by the Earth-like bulk density.

{At the very beginning of the evolution, when the planetary
atmosphere is supposed to be in boil-off, the mass-loss rates
derived from the grid are larger than those considered by
\citet{jackson2010} and \citet{leitzinger2011} by a factor of
10--10$^6$, depending on the initial planetary radius. This large
difference is caused by the fact that the energy-limited
approximation is not capable of describing atmospheric escape in
the boil-off phase. In the blow-off phase, instead, the mass-loss
rates derived from the grid are about a factor of two smaller than
those of \citet{leitzinger2011} and a factor of a few smaller than
those of \citet{jackson2010}.}

{Following the works of \citet{jackson2010} and
\citet{leitzinger2011}, we further tested the possible evolution
of CoRoT-7\,b by increasing even more the initial planetary mass
(and radius) obtaining that the planet needed to have a mass
smaller than that of Uranus (about 14.5\,\Me) to loose the primary
atmosphere within the age of the system. An initial CoRoT-7\,b
with a mass equal to that of Neptune (about 17\,\Me) would now
still hold a hydrogen-dominated envelope with $f_{\rm
at}$\,=\,0.2. We did not explore an even heavier starting point
because of the upper mass limit of 39\,\Me\ in our grid.}
\subsubsection{Results for HD219134\,b,c}
HD219134\,b,c are two close-in, transiting super-Earths orbiting a K3 main-sequence star with a radius of 0.778\,\Ro, a mass of 0.81\,\Mo, and an effective temperature of 4699\,K. The estimated age of the system is 11$\pm$2\,Gyr. The two planets have measured masses of 5.74 and 4.74\,\Me, and radii of 1.602 and 1.511\,\Re, respectively. Therefore, both planets present Earth-like densities. They orbit the host star at distances of 0.039 and 0.065\,AU, respectively.

\begin{figure}
\includegraphics[width=\hsize,clip]{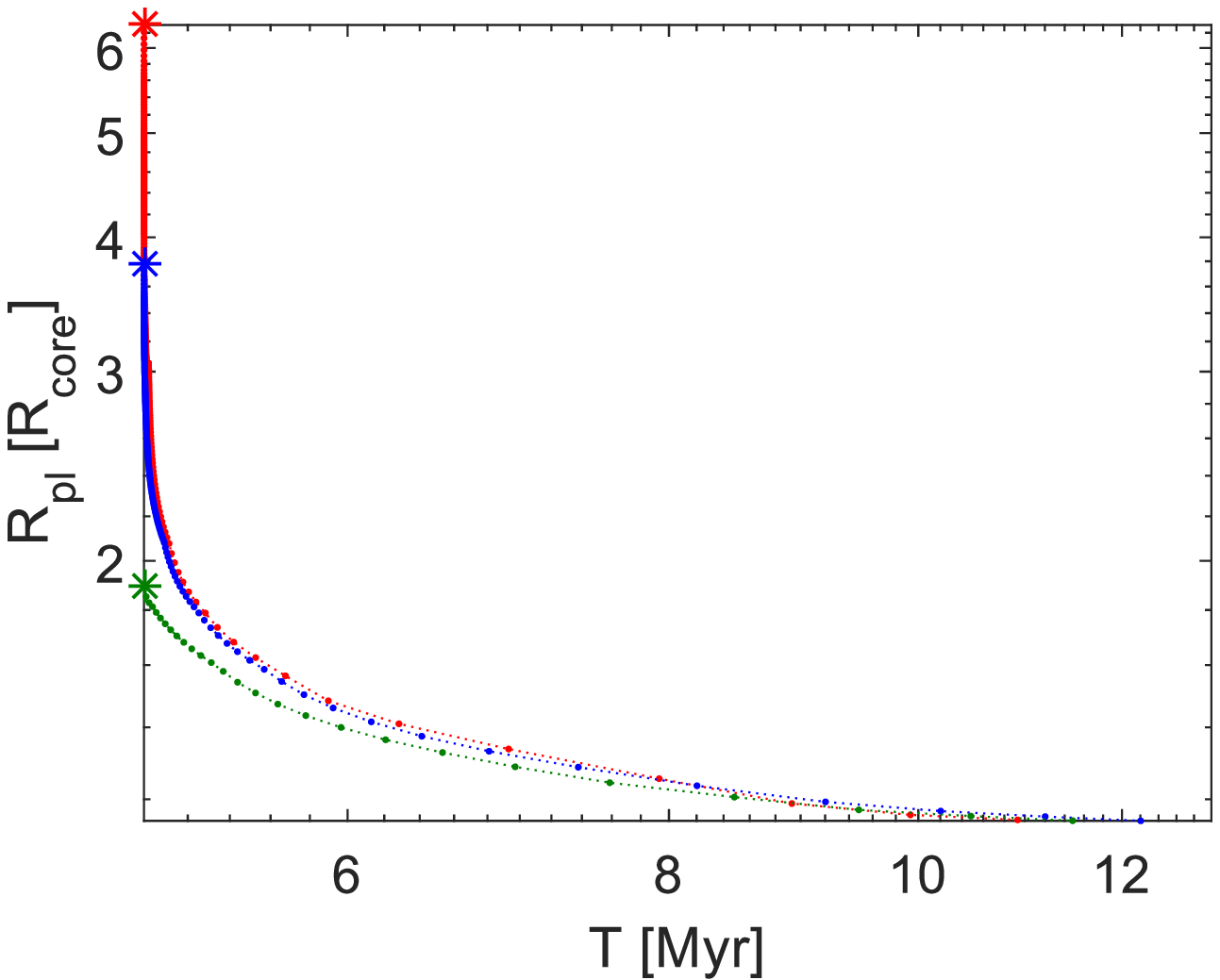}\\
\includegraphics[width=\hsize,clip]{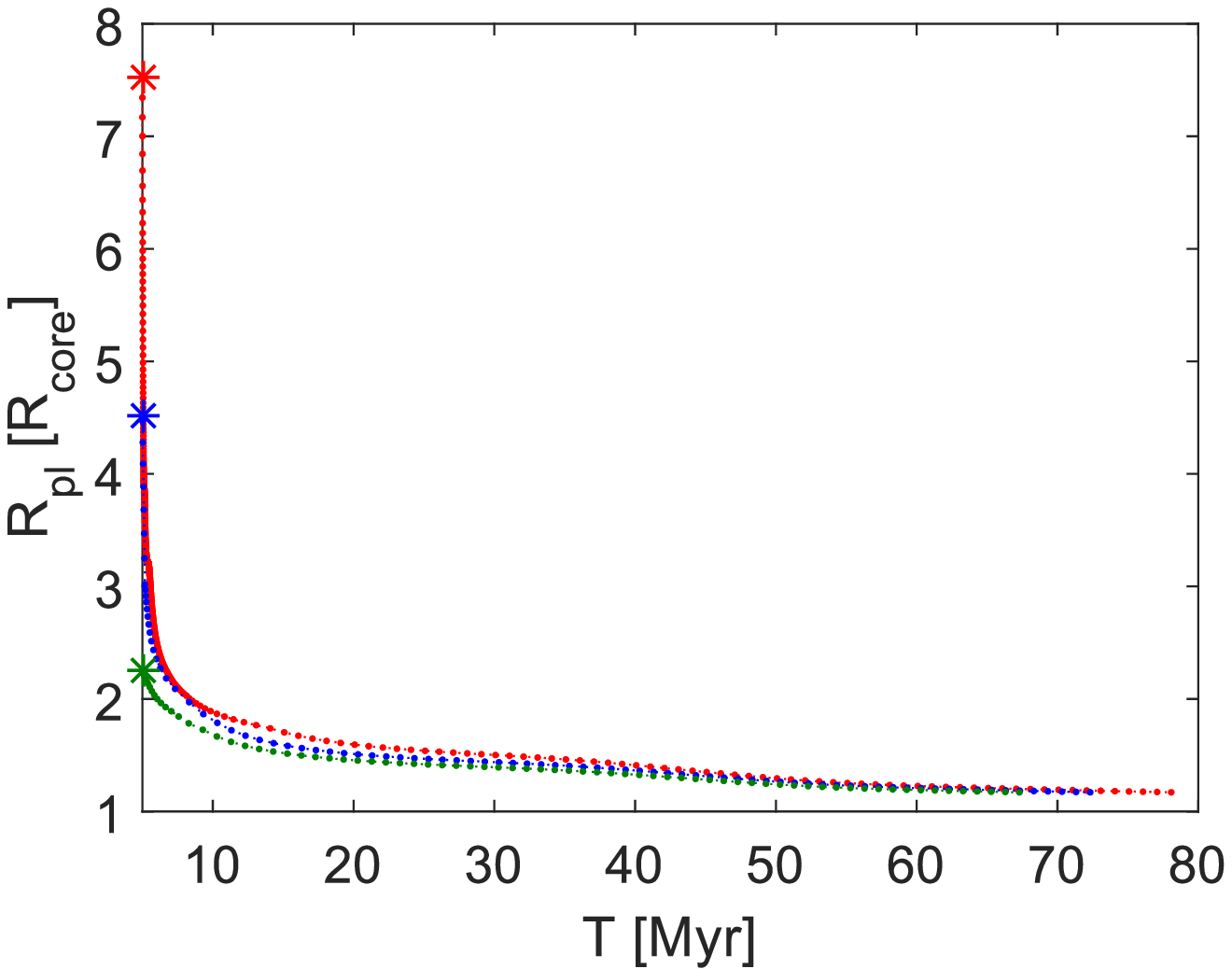}
\caption{{Evolution of the planetary radii of HD219134\,b (top)
and HD219134\,c (bottom) as a function of time. The colors,
symbols, and lines are as in Figure~\ref{fig:corot7b}.}}
\label{fig:HD219134}
\end{figure}

The evolutionary tracks, shown in Figure~\ref{fig:HD219134},
indicate that the primary, hydrogen-dominated atmosphere escaped
completely within about 12 and 80\,Myr for HD219134\,b,c,
respectively. {The difference in time between the two planets for
a complete atmospheric escape is due to their different distance
from the star, thus different equilibrium temperature and stellar
XUV irradiation.} These times are significantly shorter than the
estimated age of the system, thus allowing us to conclude that
both planets have most likely completely lost their primary,
hydrogen-dominated atmosphere through escape. This is in agreement
with \citet{dorn2018}, who arrived at the same conclusion by
employing a Bayesian inference method based on the stellar
properties and the energy-limited approximation. However, these
are extreme cases, and an approach based on the energy-limited
formulation would most likely lead to the wrong results for
younger and/or lower density planets.
%
\section{Conclusion}\label{sec:conclusion}
We upgraded and employed an existing planetary upper atmosphere
hydrodynamic code to compute a large grid of models for
{super-Earths and mini-Neptunes} orbiting late-type stars. {The
main upgrade consists in the implementation of a scheme that
automatically sets the initial parameters and profiles for each
run, thus in practice allowing one to automate computations.} The
planets covered by the grid have masses ranging from 1 to 39\,\Me\
and orbit early M- to late F-type stars in a wide range of orbital
distances, corresponding to equilibrium temperatures between 300
and 2000\,K. For each considered stellar mass, we have also
considered three different values of the XUV flux. The wide
parameter space covered by the grid allowed us to model a broad
variety of planetary atmospheres, ranging from being in boil-off,
to blow-off, and to very stable atmospheres.

For each planet in the grid, we computed the atmospheric temperature, number density, bulk velocity, X-ray and EUV volume heating rates, and abundance of the considered species as a function of distance from the planetary center. From these quantities, we estimated the positions of maximum dissociation and ionisation, the mass-loss rate, and the effective radius of the XUV absorption.

We compared the results of our grid, in particular the mass-loss rates, with those previously published for planets inside and outside the grid boundaries, finding excellent agreement. We also developed a tool to interpolate among the model results to infer the atmospheric properties of any planet covered by the grid. We took advantage of the large grid to explore in detail how the atmospheric characteristics vary with system parameters, finding for example that the mass-loss rate can be analytically described as a log-linear function of $\Lambda$ and a linear function of the stellar XUV flux.

Our grid and the interpolation routine allow one to extract in a
fraction of a second information that would otherwise require
days/weeks to obtain. This enables one to employ the results of
proper hydrodynamic computations of the mass-loss rates in
planetary atmospheric evolution calculations. This avoids the need
to use approximations, such as the energy-limited formula, that
have been shown (by various authors and further in this work) {to
significantly underestimate or overestimate in some cases} the
mass-loss rates.

We have therefore applied our grid and interpolation routine to study the evolution of the close-in, high-density planets CoRoT-7\,b and HD219134\,b,c. For CoRoT-7\,b, we found that the primary hydrogen-dominated atmosphere, assuming the planet has ever accreted one, was lost mostly through boil-off within about 0.1\,Myr. We also concluded that the planet originally could have been as massive as Uranus, because the envelope of a heavier planet would have been too massive to completely escape within the age of the system. We arrived at a similar conclusion also for HD219134\,b,c, where for these two planets we found that the hydrogen-dominated atmosphere escaped completely within about 12
and 80\,Myr, respectively. It is therefore likely that other similar planets, such as Kepler-10b and 55\,Cnc\,e, followed an analogous evolutionary path, which have left them with a secondary atmosphere formed by either outgassing from the magma ocean or sputtering of the stellar wind on the bare planetary surface, similar to what happens for Mercury \citep{mura2011,guenther2011,pfleger2015,vidotto2018}.

The simple evolutionary tracks we computed for CoRoT-7\,b and
HD219134\,b,c provide just an example of what can be achieved
using the grid and interpolation routine. There is, however, a
large number of applications in which these tools can be used and
we will explore a few of them (e.g., analytic formulation of the
mass-loss rates as a function of system parameters) in future
works currently in preparation. We are still working on increasing
the size of the grid, extending it towards more massive planets
and towards less massive stars. {The grid and interpolation
routine can be downloaded here {\tt
http://geco.oeaw.ac.at/links\_TAPAS4CHEOPS.html}. We are, however,
planning for the near future to set up a web interface allowing
users to more easily query the grid and run the interpolation
routine.}
\begin{acknowledgements}
We acknowledge the Austrian Forschungsf\"orderungsgesellschaft FFG
project ``TAPAS4CHEOPS'' P853993, the Austrian Science Fund (FWF)
NFN project S11607-N16, the FWF project P27256-N27 and the FWF
project P30949-N36. NVE acknowledges support by the RFBR grant No.
18-05-00195-a and 16-52-14006 ANF\_a. We thank the anonymous
referee for the positive approach and the useful comments that led
to a significant improvement of the manuscript.
\end{acknowledgements}
\begin{appendix}
\section{List of reactions and cross-sections employed in the model}\label{apx:A}

{The reactions and cross-sections employed in the model are
presented in Table \ref{tab:tab2}.}

\begin{table*}[t]\label{tab:tab2}
\centering \caption{Reactions and relative cross-sections employed in the model.}
\begin{tabular}{|M{4 cm}|M{6 cm}|M {4 cm}|N}
  \hline
  $\rm H \rightarrow H^+ + e$  & $\rm \nu_{H} =  5.9\times 10^{-8}\phi_{EUV} s^{-1}$ &
  \citet{storey1995}&  \\[18 pt]
  \hline
  $\rm \hh \rightarrow \hhp + e$ & $\rm \nu_{\hh} = 3.3\times 10^{-8}\phi_{EUV} s^{-1}$ & \citet{murray2009} &\\[18 pt]
  \hline
  $\rm H^{+} + e \rightarrow H$ & $\rm \alpha_{H} = 4\times 10^{-12}(300/T)^{0.64} cm^{3}s^{-1}$ & \citet{yelle2004}&\\[18 pt]
  \hline
  $\rm \hhp + e \rightarrow H + H$ & $\rm \alpha_{\hh} = 2.3\times 10^{-8} (300/T)^{0.4} cm^3 s^{-1}$ & \citet{yelle2004} &\\[18 pt]
  \hline
  $\rm \hh \rightarrow H + H$ & $\rm \nu_{diss} = 1.5\times 10^{-9} e^{(-49000/T)} cm^{3}s^{-1}$ & \citet{yelle2004} &\\[18 pt]
  \hline
  $\rm H + H \rightarrow \hh$ & $\rm \gamma_H = 8.0\times 10^{-33} (300/T)^{0.6} cm^{3}s^{-1}$ & \citet{yelle2004} &\\[18 pt]
  \hline
  $\rm H + e \rightarrow H^+$ & $\rm \nu_{Hcol} = 5.9\times 10^{-11}T^{1/2}e^{(-157809/T)} cm^{3}s^{-1}$ & \citet{black1981} &\\[18 pt]
  \hline
  $\rm \hhp + \hh \rightarrow \hhh + H$ & $\rm \gamma_{\hh} = 2\times 10^{-9} cm^{3}s^{-1}$ & \citet{yelle2004} &\\[18 pt]
  \hline
  $\rm \hhh + H \rightarrow \hhp + \hh$ & $\rm \gamma_{\hh} = 2\times 10^{-9} cm^{3}s^{-1}$ & \citet{yelle2004} &\\[18 pt]
  \hline
  $\rm \hhh + e \rightarrow \hh + H$ & $\rm \alpha_{\hhh1} = 2.9\times 10^{-8}(\frac{300}{T_e})^{0.65} cm^{3}s^{-1}$ & \citet{yelle2004} &\\[18 pt]
  \hline
  $\rm \hhh + e \rightarrow H + H + H$ & $\rm \alpha_{\hhh2} = 8.6\times 10^{-8}(\frac{300}{T_e})^{0.65} cm^{3}s^{-1}$ & \citet{yelle2004} &\\[18 pt]
  \hline
\end{tabular}
\end{table*}

%
\section{Normalisations employed for the computation of each model}\label{apx:B}
Here we give the normalisations used for the computation of the models.
\begin{eqnarray*}
\tilde{r} &=& r/R_{\rm pl}\,, \\
\tilde{T} &=& T/T_{\rm eq}\,, \\
\tilde{\rho} &=& \rho/\rho_{0}~~{\rm where}~~\rho_{0} = N_{0}m_{\rm H_2}\,, \\
N_{0} &=& P_0/(2kT_{\rm eq})\,, \\
\tilde{V} &=& V/C_{\rm s0}~~{\rm where}~~C_{\rm s0} = \sqrt{kT_{\rm eq}/m_{\rm H}}\,, \\
\tilde{U} &=& m_{\rm H}U/(kT_{\rm eq})\,, \\
\tilde{P} &=& P/P_0\,, \\
X &=& m_{\rm H}n_{\rm H}/\rho~~{\rm and}~~X^+ = m_{\rm H^+}n_{\rm H^+}/\rho\,, \\
Y &=& m_{\rm H_2}n_{\rm H_2}/\rho~~{\rm and}~~Y^+ = m_{\rm H_2^+}n_{\rm H_2^+}/\rho\,, \\
Z^+ &=& n_{\rm H_3^+}m_{\rm H_3^+}/\rho\,, \\
\tilde{Q}_{\rm m} &=& \eta_{\rm m}\phi_{\rm m}R_{\rm pl}/(m_{\rm H_2}C_{\rm s0}^3)~~{\rm where}~~m = {\rm X, EUV}\,, \\
\tilde{Q}_{\rm Ly\alpha} &=& 7.5\times10^{-19}N_0R_{\rm pl}/(m_{\rm H_2}C_{\rm s0}^3), \\
\tilde{\nu}_{\rm H} &=& \nu_{\rm H}R_{\rm pl}/C_{\rm s0}~~{\rm
and}~~\tilde{\nu}_{\rm H_2} =
\nu_{\rm H_2}R_{\rm pl}/C_{\rm s0}\,, \\
\tilde{\alpha}_{\rm H} &=& \alpha_{\rm H}N_0R_{\rm pl}/C_{\rm s0}~~{\rm and}~~\tilde{\alpha}_{\rm H_2} = \alpha_{\rm H_2}N_0R_{\rm pl}/C_{\rm s0}\,, \\
\tilde{\nu}_{\rm Hcol} &=& \nu_{\rm Hcol}N_0R_{\rm pl}/C_{\rm s0}\,, \\
\tilde{\nu}_{\rm diss} &=& \nu_{\rm diss}N_0R_{\rm pl}/C_{\rm s0}\,, \\
\tilde{\gamma}_{\rm H} &=& \gamma_{\rm H}N_0^2R_{\rm pl}/C_{\rm s0}\,, \\
\tilde{\chi} &=& \chi T_{\rm eq}/(\rho_0R_{\rm pl}C_{\rm s0}^3)\,.
\end{eqnarray*}
In the equations above, the subscript ``0'' denotes the values at the lower boundary, e.g., $P_0$ and $C_{s0}$ are respectively the pressure and sound speed at the lower boundary.
\section{Planetary atmospheric mass-loss rates as a function of system parameters}\label{apx:C}
\begin{figure*}[ht!]
\begin{rotate}{90}\hspace{2.3 cm} $d_0 = 0.0608$\,au \hspace{2.3 cm} $d_0 = 0.1131$\,au \hspace{2.3 cm} $d_0 = 0.2792$\,au \end{rotate}
\includegraphics[width=\hsize,clip]{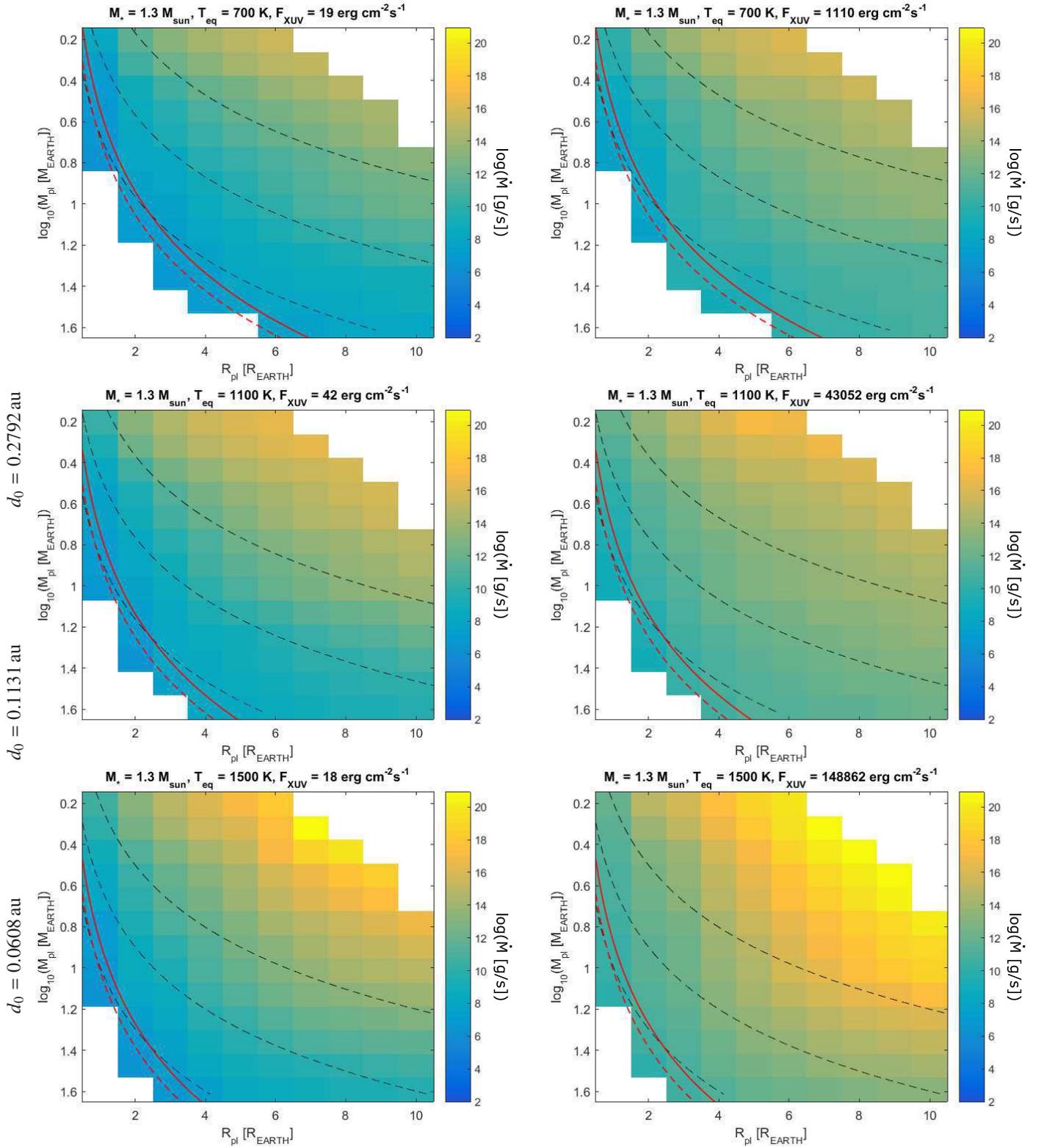}
\caption{Same as Figure~\ref{fig:escapes}, but for a stellar mass
of 1.3\,\Mo.} \label{fig:escapes13}
\end{figure*}

\begin{figure*}[ht!]
\begin{rotate}{90}\hspace{2.3 cm} $d_0 = 0.019$\,au \hspace{2.3 cm} $d_0 = 0.0361$\,au \hspace{2.3 cm} $d_0 = 0.089$\,au \end{rotate}
\includegraphics[width=\hsize,clip]{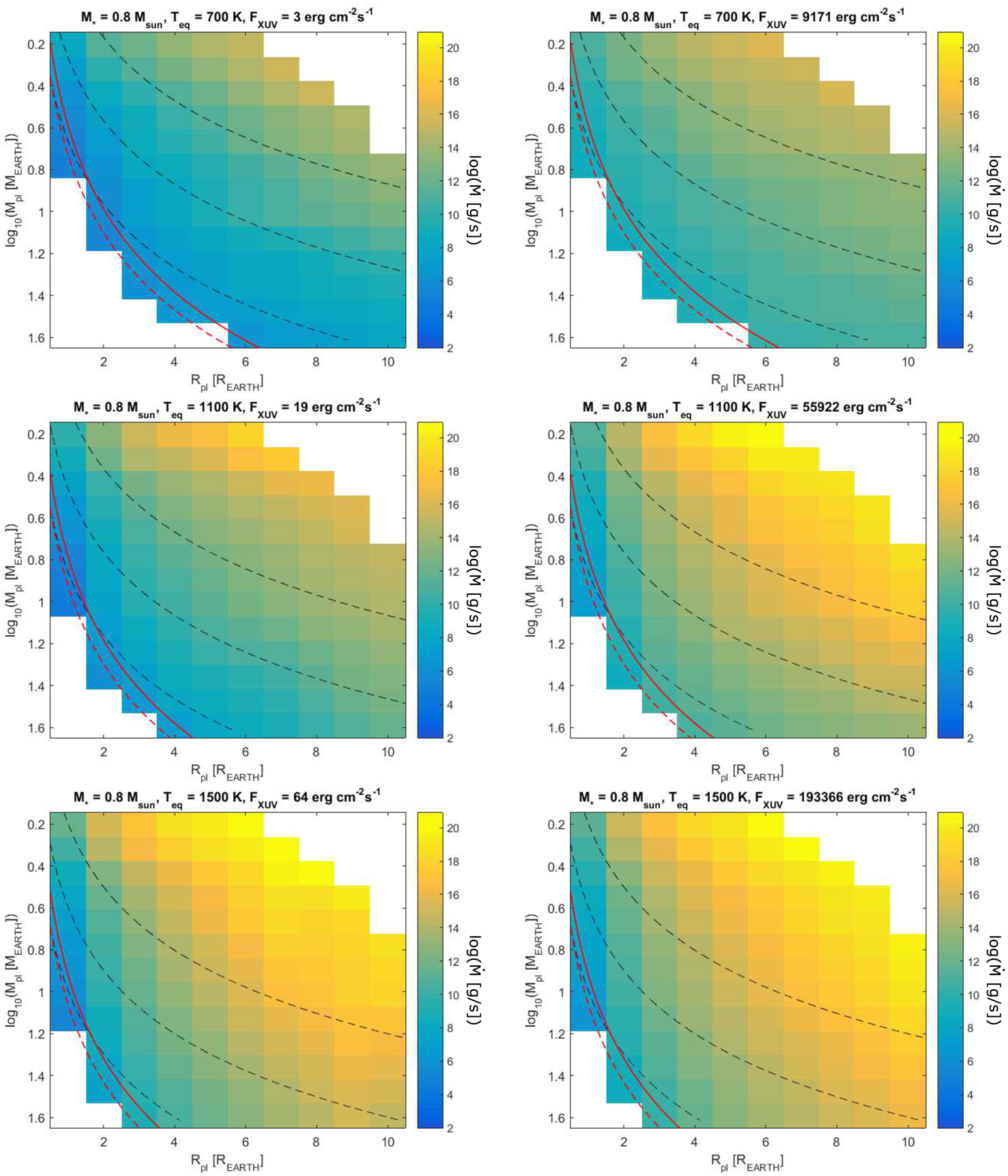}
\caption{Same as Figure~\ref{fig:escapes}, but for a stellar mass
of 0.8\,\Mo.} \label{fig:escapes08}
\end{figure*}

\begin{figure*}[ht!]
\begin{rotate}{90}\hspace{2.3 cm} $d_0 = 0.0097$\,au \hspace{2.3 cm} $d_0 = 0.018$\,au \hspace{2.3 cm} $d_0 = 0.0445$\,au \end{rotate}
\includegraphics[width=\hsize,clip]{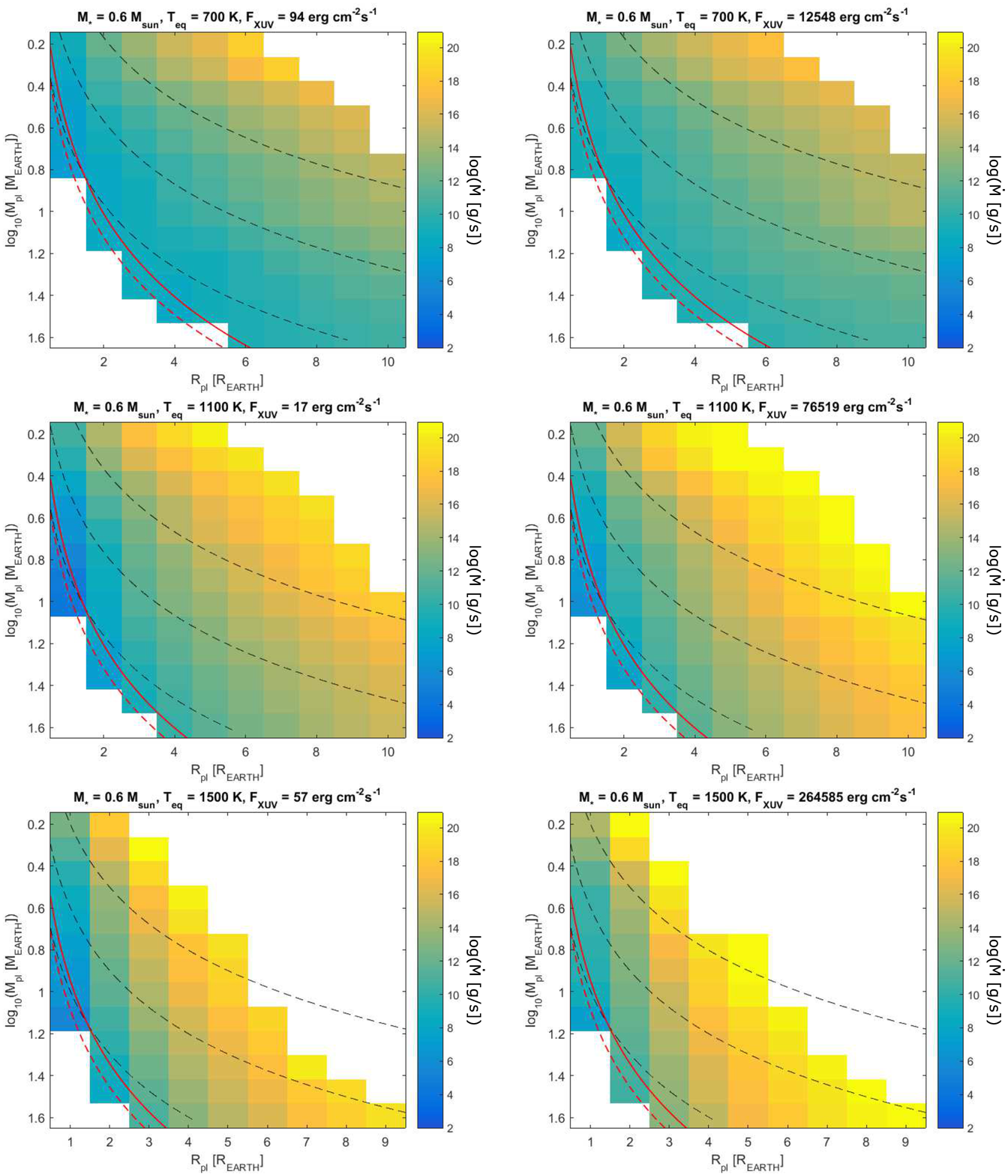}
\caption{Same as Figure~\ref{fig:escapes}, but for a stellar mass
of 0.6\,\Mo.} \label{fig:escapes06}
\end{figure*}

\begin{figure*}[ht!]
\begin{rotate}{90}\hspace{2.3 cm} $d_0 = 0.0069$\,au \hspace{2.3 cm} $d_0 = 0.017$\,au \hspace{2.3 cm} $d_0 = 0.0926$\,au \end{rotate}
\includegraphics[width=\hsize,clip]{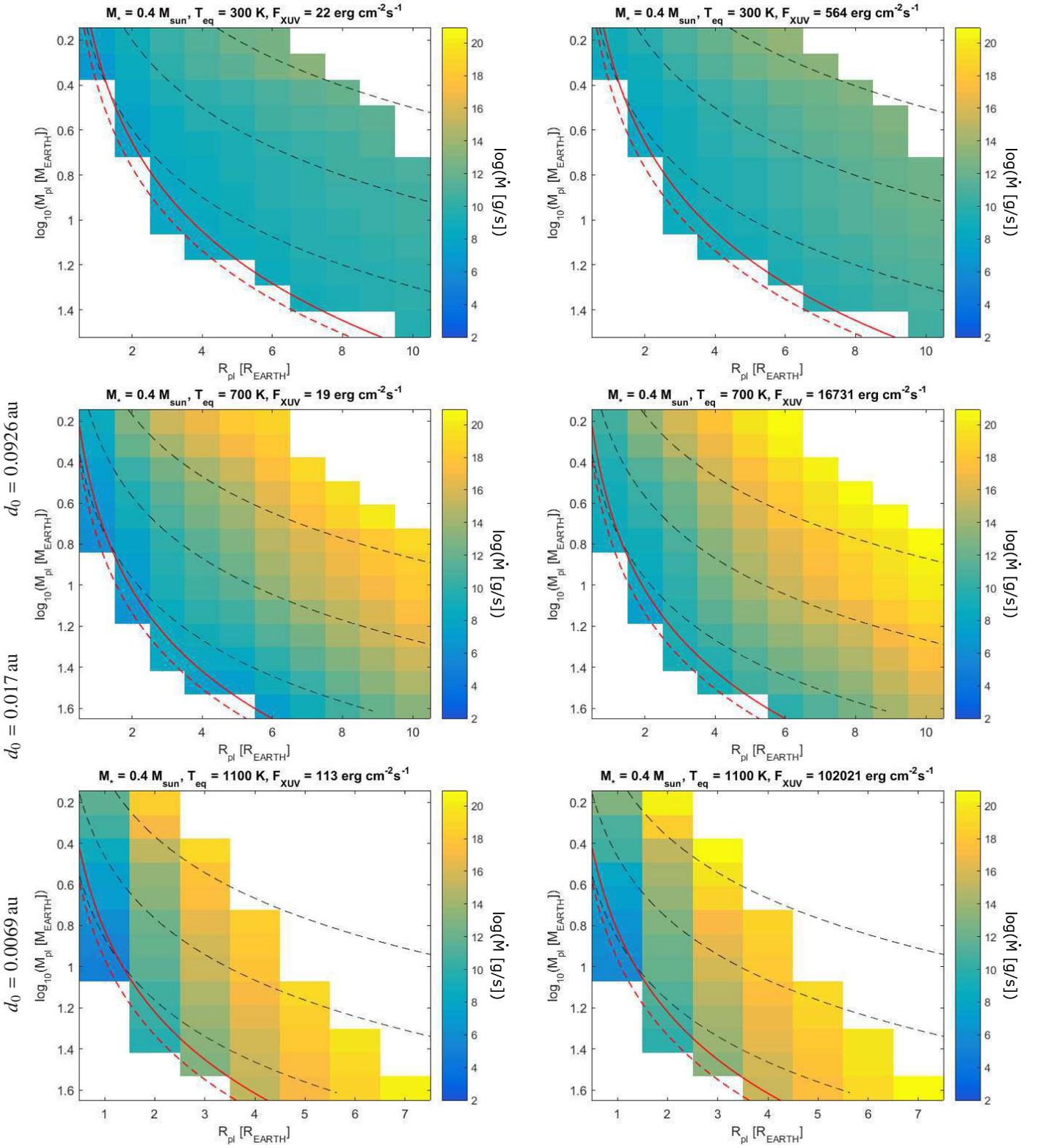}
\caption{{Same as Figure~\ref{fig:escapes}, but for a stellar mass
of 0.4\,\Mo. In these plots, the temperature range was shifted to
$300-1100$\,K, instead of $700-1500$\,K, to make the range of
orbital separations comparable to that of other stellar masses and
because of the cut on the Roche lobe for planets with an
equilibrium temperature of 1500\,K orbiting 0.4\,\Mo\ stars.}}
\label{fig:escapes04}
\end{figure*}

We present here plots analogous to those in Figure~\ref{fig:escapes}, but for different stellar masses (Figures~\ref{fig:escapes13} to \ref{fig:escapes04}).

%
\end{appendix}
%
%
%
%
%

\begin{thebibliography}{}
\bibitem[Barros et al.(2014)]{barros2014}
    Barros, S.~C.~C., Almenara, J.~M., Deleuil, M., et al.\ 2014,
\aap, 569, A74
\bibitem[Barclay et al.(2018)]{barclay2018}
    Barclay, T., Pepper, J., \& Quintana, E.~V.\ 2018, AJ, submitted (arXiv:1804.05050)
\bibitem[Batygin \& Stevenson(2013)]{batygin2013}
    Batygin, K., \& Stevenson, D.~J.\ 2013, \apjl, 769, L9
\bibitem[Black(1981)]{black1981}
    Black, J.~H.\ 1981, \mnras, 197, 553,
\bibitem[Bonfils et al.(2013)]{bonfils2013}
    Bonfils, X., Delfosse, X., Udry, S., et al.\ 2013, \aap, 549, A109
\bibitem[Bourrier \& Lecavelier des Etangs(2013)]{bourrier2013}
    Bourrier, V., \& Lecavelier des Etangs, A.\ 2013, \aap, 557, A124
\bibitem[Bourrier et al.(2016)]{bourrier2016}
    Bourrier, V., Lecavelier des Etangs, A., Ehrenreich, D., Tanaka, Y.~A.,         \& Vidotto, A.~A.\ 2016, \aap, 591, A121
\bibitem[Broeg et al.(2013)]{broeg2013}
    Broeg, C., Fortier, A., Ehrenreich, D., et al.\ 2013, European Physical         Journal Web of Conferences, 47, 03005
\bibitem[Cecchi-Pestellini et al.(2009)]{cecchi2009}
    Cecchi-Pestellini, C., Ciaravella, A., Micela, G., \& Penz, T.\ 2009,           \aap, 496, 863
\bibitem[Chamberlain(1963)]{chamber1963}
    Chamberlain, J.~W.\ 1963, \planss, 11, 901
\bibitem[Chen \& Rogers(2016)]{chen2016}
    Chen, H., \& Rogers, L.~A.\ 2016, \apj, 831, 180
\bibitem[Cubillos et al.(2017a)]{cubillos2017a}
    Cubillos, P., Erkaev, N.~V., Juvan, I., et al.\ 2017a, \mnras, 466, 1868
\bibitem[Cubillos et al.(2017b)]{cubillos2017b}
    Cubillos, P.~E., Fossati, L., Erkaev, N.~V., et al.\ 2017b, \apj, 849,          145
\bibitem[Deming et al.(2009)]{deming2009}
    Deming, D., Seager, S., Winn, J., et al.\ 2009, \pasp, 121, 952
\bibitem[Dorn \& Heng(2018)]{dorn2018}
    Dorn, C., \& Heng, K.\ 2018, \apj, 853, 64
\bibitem[Ehrenreich et al.(2008)]{ehrenreich2008}
    Ehrenreich, D., Lecavelier Des Etangs, A., H{\'e}brard, G., et al.\             2008, \aap, 483, 933
\bibitem[Ehrenreich and D\'{e}sert(2011)]{ehrenreich2011}
    Ehrenreich and D\'{e}sert \aap, 2011
\bibitem[Ehrenreich et al.(2015)]{ehrenreich2015}
    Ehrenreich, D., Bourrier, V., Wheatley, P.~J., et al.\ 2015, \nat, 522,         459
\bibitem[Erkaev et al.(2007)]{erkaev2007}
    Erkaev, N.~V., Kulikov, Y.~N., Lammer, H., et al.\ 2007, \aap, 472, 329
\bibitem[Erkaev et al.(2013)]{erkaev2013}
    Erkaev, N.~V., Lammer, H., Odert, P., et al.\ 2013, Astrobiology, 13,           1011
\bibitem[Erkaev et al.(2014)]{erkaev2014}
    Erkaev, N.~V., Lammer, H., Elkins-Tanton, L.~T., et al.\ 2014, \planss,         98, 106
\bibitem[Erkaev et al.(2015)]{erkaev2015}
    Erkaev, N.~V., Lammer, H., Odert, P., Kulikov, Y.~N., \& Kislyakova,            K.~G.\ 2015, \mnras, 448, 1916
\bibitem[Erkaev et al.(2016)]{erkaev2016}
    Erkaev, N.~V., Lammer, H., Odert, P., et al.\ 2016, \mnras, 460, 1300
\bibitem[Erkaev et al.(2017)]{erkaev2017}
    Erkaev, N.~V., Odert, P., Lammer, H., et al.\ 2017, \mnras, 470, 4330
\bibitem[Fleming et al.(2018)]{fleming2018}
    Fleming, B.~T., France, K., Nell, N., et al.\ 2018, Journal of Astronomical Telescopes, Instruments, and Systems, 4, 014004
\bibitem[Fossati et al.(2015)]{fossati2015}
    Fossati, L., France, K., Koskinen, T., et al.\ 2015, \apj, 815, 118
\bibitem[Fossati et al.(2017)]{fossati2017}
    Fossati, L., Erkaev, N.~V., Lammer, H., et al.\ 2017, \aap, 598, A90
\bibitem[Fossati et al.(2018)]{fossati2018}
    Fossati, L., Koskinen, T., France, K., et al.\ 2018, \aj, 155, 113
\bibitem[Fulton et al.(2017)]{fulton2017}
    Fulton, B.~J., Petigura, E.~A., Howard, A.~W., et al.\ 2017, \aj, 154,          109
\bibitem[Fulton \& Petigura(2018)]{fulton2018}
    Fulton, B.~J., \& Petigura, E.~A.\ 2018, \aj, submitted (arXiv:1805.01453)
\bibitem[Gardner et al.(2006)]{gardner2006}
    Gardner, J.~P., Mather, J.~C., Clampin, M., et al.\ 2006, \ssr, 123, 485
\bibitem[Gillon et al.(2017)]{gillon2017}
    Gillon, M., Demory, B.-O., Van Grootel, V., et al.\ 2017, Nature Astronomy, 1, 0056
\bibitem[Ginzburg et al.(2016)]{ginzburg2016}
    Ginzburg, S., Schlichting, H.~E., \& Sari, R.\ 2016, \apj, 825, 29
\bibitem[Ginzburg et al.(2018)]{ginzburg2018}
    Ginzburg, S., Schlichting, H.~E., \& Sari, R.\ 2018, \mnras, 476, 759
\bibitem[Guenther et al.(2011)]{guenther2011}
    Guenther, E.~W., Cabrera, J., Erikson, A., et al.\ 2011, \aap, 525, A24
\bibitem[Guo \& Ben-Jaffel(2016)]{guo2016}
    Guo, J.~H., \& Ben-Jaffel, L.\ 2016, \apj, 818, 107
\bibitem[Howe et al.(2014)]{howe2014}
    Howe, A.~R., Burrows, A., \& Verne, W.\ 2014, \apj, 787, 173
\bibitem[Jackson et al.(2010)]{jackson2010}
    Jackson, B., Miller, N., Barnes, R., et al.\ 2010, \mnras, 407, 910
\bibitem[Jackson et al.(2012)]{jackson2012}
    Jackson, A.~P., Davis, T.~A., \& Wheatley, P.~J.\ 2012, \mnras, 422,            2024
\bibitem[Jeans(1925)]{jeans1925}
    Jeans, J.\ 1925, The Dynamical Theory of Gases. By Sir James Jeans.             Cambridge University Press, 1925. ISBN: 978-1-1080-0564-7
\bibitem[Jin et al.(2014)]{jin2014}
    Jin, S., Mordasini, C., Parmentier, V., et al.\ 2014, \apj, 795, 65
\bibitem[Jin \& Mordasini(2017)]{jin2017}
    Jin, S., \& Mordasini, C.\ 2018, \apj, 853, 163
\bibitem[Johnstone et al.(2015)]{johnstone2015}
    Johnstone, C.~P., G{\"u}del, M., St{\"o}kl, A., et al.\ 2015, \apjl, 815, L12
\bibitem[Johnstone et al.(2015b)]{johnstone2015rot}
    Johnstone, C.~P., G{\"u}del, M., Brott, I., \& L{\"u}ftinger, T.\ 2015, \aap, 577, A28
\bibitem[Kislyakova et al.(2014)]{kislyakova2014}
    Kislyakova, K.~G., Johnstone, C.~P., Odert, P., et al.\ 2014, \aap, 562,    A116
\bibitem[Koskinen et al.(2010)]{koskinen2010}
    Koskinen, T.~T., Yelle, R.~V., Lavvas, P., \& Lewis, N.~K.\ 2010, \apj,         723, 116
\bibitem[Koskinen et al.(2013)]{koskinen2013}
    Koskinen, T.~T., Harris, M.~J., Yelle, R.~V., \& Lavvas, P.\ 2013, \icarus, 226, 1678
\bibitem[Kubyshkina et al.(2018)]{kubyshkina2018}
    Kubyshkina, D., Lendl, M., Fossati, L., et al.\ 2018, \aap, 612, A25
\bibitem[Lammer et al.(2003)]{lammer2003}
    Lammer, H., Selsis, F., Ribas, I., et al.\ 2003, \apjl, 598, L121
\bibitem[Lammer et al.(2009)]{lammer2009}
    Lammer, H., Odert, P., Leitzinger, M., et al.\ 2009, \aap, 506, 399
\bibitem[Lammer et al.(2013)]{lammer2013}
    Lammer, H., Erkaev, N.~V., Odert, P., et al.\ 2013, \mnras, 430, 1247
\bibitem[Lammer et al.(2016)]{lammer2016}
    Lammer, H., Erkaev, N.~V., Fossati, L., et al.\ 2016, \mnras, 461, L62
\bibitem[Lecavelier des Etangs et al.(2004)]{Lecavelier2004}
    Lecavelier des Etangs et al.  418, L1, \aap 2004
\bibitem[Leitzinger et al.(2011)]{leitzinger2011}
    Leitzinger, M., Odert, P., Kulikov, Y.~N., et al.\ 2011, \planss, 59, 1472
\bibitem[L{\'e}ger et al.(2009)]{leger2009}
    L{\'e}ger, A., Rouan, D., Schneider, J., et al.\ 2009, \aap, 506, 287
\bibitem[Lopez \& Fortney(2013)]{lopez2013}
    Lopez, E.~D., \& Fortney, J.~J.\ 2013, \apj, 776, 2
\bibitem[Lopez \& Fortney(2014)]{lopez2014}
    Lopez, E.~D., \& Fortney, J.~J.\ 2014, \apj, 792, 1
\bibitem[Lopez(2017)]{lopez2017}
    Lopez, E.~D.\ 2017, \mnras, 472, 245
\bibitem[Lundkvist et al.(2016)]{lundkvist2016}
    Lundkvist, M.~S., Kjeldsen, H., Albrecht, S., et al.\ 2016, Nature          Communications, 7, 11201
\bibitem[Mamajek \& Hillenbrand(2008)]{mamajek2008}
    Mamajek, E.~E., \& Hillenbrand, L.~A.\ 2008, \apj, 687, 1264-1293
\bibitem[Mamajek(2009)]{mamajek2009}
    Mamajek, E.~E.\ 2009, American Institute of Physics Conference Series, 1158, 3
\bibitem[Masuda(2014)]{masuda2014}
    Masuda, K.\ 2014, \apj, 783, 53
\bibitem[Menager et al.(2013)]{menager2013}
    Menager, H., Barth{\'e}lemy, M., Koskinen, T., et al.\ 2013, \icarus, 226, 1709
\bibitem[Miller et al.(2013)]{miller2013}
    Miller, S., Stallard, T., Tennyson, J., \& Melin, H.\ 2013, Journal of          Physical Chemistry A, 117, 9770
\bibitem[Motalebi et al.(2015)]{motalebi2015}
    Motalebi, F., Udry, S., Gillon, M., et al.\ 2015, \aap, 584, A72
\bibitem[Mullally et al.(2015)]{mullally2015}
    Mullally, F., Coughlin, J.~L., Thompson, S.~E., et al.\ 2015, \apjs,            217, 31
\bibitem[Mura et al.(2011)]{mura2011}
    Mura, A., Wurz, P., Schneider, J., et al.\ 2011, \icarus, 211, 1
\bibitem[Murray-Clay et al.(2009)]{murray2009}
    Murray-Clay, R.~A., Chiang, E.~I., \& Murray, N.\ 2009, \apj, 693, 23
\bibitem[{\"O}pik(1963)]{opik1963}
    {\"O}pik, E.~J.\ 1963, Geophysical Journal, 7, 490
\bibitem[Owen \& Jackson(2012)]{owen2012}
    Owen, J.~E., \& Jackson, A.~P.\ 2012, \mnras, 425, 2931
\bibitem[Owen \& Mohanty(2016)]{owen2016a}
    Owen, J.~E., \& Mohanty, S.\ 2016, \mnras, 459, 4088
\bibitem[Owen \& Wu(2016)]{owen2016b}
    Owen, J.~E., \& Wu, Y.\ 2016, \apj, 817, 107
\bibitem[Owen \& Wu(2017)]{owen2017}
    Owen, J.~E., \& Wu, Y.\ 2017, \apj, 847, 29
\bibitem[Paxton et al.(2018)]{paxton2018}Paxton, B., Schwab, J., Bauer, E.~B., et al.\ 2018, \apjs, 234, 34
\bibitem[Pfleger et al.(2015)]{pfleger2015}
    Pfleger, M., Lichtenegger, H.~I.~M., Wurz, P., et al.\ 2015, \planss, 115, 90
\bibitem[Pizzolato et al.(2003)]{pizzolato2003}
    Pizzolato, N., Maggio, A., Micela, G., Sciortino, S., \& Ventura, P.\ 2003, \aap, 397, 147
\bibitem[Quirrenbach et al.(2010)]{carmenes}
    Quirrenbach, A., Amado, P.~J., Mandel, H., et al.\ 2010, \procspie,             7735, 773513
\bibitem[Rauer et al.(2014)]{rauer2014}
    Rauer, H., Catala, C., Aerts, C., et al.\ 2014, Experimental Astronomy,         38, 249
\bibitem[Reiners et al.(2014)]{reiners2014}
    Reiners, A., Sch{\"u}ssler, M., \& Passegger, V.~M.\ 2014, \apj, 794,           144
\bibitem[Ribas et al.(2005)]{ribas2005}
    Ribas, I., Guinan, E.~F., G{\"u}del, M., \& Audard, M.\ 2005, \apj, 622, 680
\bibitem[Ricker et al.(2015)]{ricker2015}
    Ricker, G.~R., Winn, J.~N., Vanderspek, R., et al.\ 2015, Journal of Astronomical Telescopes, Instruments, and Systems, 1, 014003
\bibitem[Rodr{\'{\i}}guez et al.(2016)]{rodrigues2016}
    Rodr{\'{\i}}guez, A., Callegari, N., \& Correia, A.~C.~M.\ 2016, \mnras, 463, 3249
\bibitem[Salz et al.(2016)]{salz2016}
    Salz, M., Schneider, P.~C., Czesla, S., \& Schmitt, J.~H.~M.~M.\ 2016,          \aap, 585, L2
\bibitem[Sanz-Forcada et al.(2011)]{sanz2011}
    Sanz-Forcada, J., Micela, G., Ribas, I., et al.\ 2011, \aap, 532, A6
\bibitem[Shematovich et al.(2014)]{shematovich2014}
    Shematovich, V.~I., Ionov, D.~E., \& Lammer, H.\ 2014, \aap, 571, A94
\bibitem[Spitzer(1978)]{spitzer1978}
    Spitzer, L.\ 1978, Physical processes in the interstellar medium, by Lyman Spitzer. New York Wiley-Interscience, 1978, 333
\bibitem[Storey \& Hummer(1995)]{storey1995}
     Storey, P.~J., \& Hummer, D.~G.\ 1995, \mnras, 272(1), 41-48
\bibitem[St{\"o}kl et al.(2015)]{stokl2015}
    St{\"o}kl, A., Dorfi, E., \& Lammer, H.\ 2015, \aap, 576, A87
\bibitem[St{\"o}kl et al.(2016)]{stokl2016}
    St{\"o}kl, A., Dorfi, E.~A., Johnstone, C.~P., \& Lammer, H.\ 2016, \apj, 825, 86
\bibitem[Tinetti et al.(2017)]{tinetti2017}
    Tinetti, G., Drossart, P., Eccleston, P., et al.\ 2017, European Planetary Science Congress, 11, EPSC2017-713
\bibitem[Tu et al.(2015)]{tu2015}
    Tu, L., Johnstone, C.~P., G{\"u}del, M., \& Lammer, H.\ 2015, \aap, 577, L3
\bibitem[Valencia et al.(2010)]{valencia2010}
    Valencia, D., Ikoma, M., Guillot, T., \& Nettelmann, N.\ 2010, \aap, 516, A20
\bibitem[Van Eylen et al.(2017)]{vaneylen2018}
    Van Eylen, V., Agentoft, C., Lundkvist, M.~S., et al.\ 2018, \mnras, 479, 4786
\bibitem[Vidotto et al.(2018)]{vidotto2018}
    Vidotto, A.~A., Lichtenegger, H., Fossati, L., et al.\ 2018, \mnras, in press (arXiv:1808.00404)
\bibitem[Vogt et al.(2015)]{vogt2015}
    Vogt, S.~S., Burt, J., Meschiari, S., et al.\ 2015, \apj, 814, 12
\bibitem[Watson et al.(1981)]{watson1981}
    Watson, A.~J., Donahue, T.~M., \& Walker, J.~C.~G.\ 1981, \icarus, 48,          150
\bibitem[Weiss \& Marcy(2014)]{weiss2014}
    Weiss, L.~M., \& Marcy, G.~W.\ 2014, \apjl, 783, L6
\bibitem[Wolfgang et al.(2016)]{wolfgang2016}
    Wolfgang, A., Rogers, L.~A., \& Ford, E.~B.\ 2016, \apj, 825, 19
\bibitem[Wright et al.(2011)]{wright2011}
    Wright, N.~J., Drake, J.~J., Mamajek, E.~E., \& Henry, G.~W.\ 2011,             \apj, 743, 48
\bibitem[Yelle(2004)]{yelle2004}
    Yelle, R.~V.\ 2004, \icarus, 170, 167
\bibitem[Yi et al.(2001)]{yi2001}
    Yi, S., Demarque, P., Kim, Y.-C., et al.\ 2001, \apjs, 136, 417
\end{thebibliography}
\end{document}